\documentclass[journal]{IEEEtran}
\ifCLASSINFOpdf
\else
\fi

\usepackage{graphicx}
\usepackage{mathptmx} 
\usepackage{times} 
\usepackage{amsmath} 
\usepackage{amssymb}  
\usepackage{physics}
\usepackage{color}
\usepackage{physics}
\usepackage{booktabs}
\usepackage{breqn}
\usepackage{siunitx}
\usepackage{subcaption}
\usepackage{algorithm,algorithmic}
\usepackage{nomencl}
\usepackage{multirow}
\usepackage{mathtools}
\usepackage{booktabs}
\makenomenclature

\newtheorem{prob}{Problem}

\begin{document}
%
\title{Information-Driven Path Planning for UAV with Limited Autonomy in Large-scale Field Monitoring}
%
%
%

\author{Nicolas Bono Rossello$^{1,2}$ \and Renzo Fabrizio Carpio$^{2}$ \and Andrea Gasparri$^{2}$ \and  Emanuele Garone$^{1}$
\thanks{This work has been supported by the European Commission under the Grant Agreement number 774571(project PANTHEON- ``Precision farming of hazelnut orchards")}
\thanks{$^{1}$Service d’Automatique et d’Analyse des Systèmes: Université Libre
de Bruxelles (ULB), Av. F.D. Roosvelt 50, CP 165/55, 1050 - Brussels,
Belgium. Email: 
        {\tt\small \{nbonoros,emgarone\}@ulb.ac.be}}%
\thanks{$^{2}$Department of  Engineering: Roma Tre University,
        Via della Vasca Navale, 79/81 00146 - Rome, Italy. Email: 
        {\tt\small renzo.carpio@uniroma3.it}, 
        {\tt\small gasparri@inf.uniroma3.it}}%
}

%
%

\markboth{}%
{Shell \MakeLowercase{\textit{et al.}}: Bare Demo of IEEEtran.cls for IEEE Journals}
%



\maketitle

\begin{abstract}
This paper presents a novel information-based mission planner for a drone tasked to monitor a spatially distributed dynamical phenomenon. For the sake of simplicity, the area to be monitored is discretized. 
The insight behind the proposed approach is that, thanks to the spatio-temporal dependencies of the observed phenomenon, one does not need to collect data on the entire area, which is one of the main limiting factors in UAV applications due to their limited autonomy. In fact, unmeasured states can be estimated using an estimator, such as a Kalman filter. In this context the planning problem becomes the one of generating a flight plan that maximizes the quality of the state estimation while satisfying the flight constraints (e.g. flight time). 
The first result of this paper is the formulation of this problem as a special Orienteering Problem where the cost function is a measure of the quality of the estimation. This results in a Mixed-Integer Semi-Definite formulation which can be optimally solved for small instances of the problem. For larger instances, a heuristic is proposed which provides sub-optimal results. Simulations numerically demonstrate the capabilities and efficiency of the proposed path planning strategy. We believe this approach has the potential to increase dramatically the area that a drone can monitor, thus increasing the number of applications where monitoring with drones can become economically convenient. 
\end{abstract}

Note to Practitioners:
\begin{abstract}
This paper was motivated by the problem of performing large-scale field monitoring activities using UAVs, which at the moment is very time consuming and limits the definitive adoption of UAVs for this kind of activities. This problem is caused by the limited autonomy of commercial UAVs and the lack of systematic ways to plan missions so as to maximize the amount of information collected. This work starts from the observation that, in many applications, the phenomena that one wants to observe  have dynamics and statistical properties. Accordingly, data that are not directly measured can be estimated with a characterizable observation error. In this paper, we develop the theoretical foundations for an information-based path planning and define the problem of designing the optimal mission as an optimization problem based on the knowledge of the monitored phenomenon. The presented results have the potential to dramatically improve the effectiveness of drones for monitoring applications. 
\end{abstract}


%
\IEEEpeerreviewmaketitle

\section{Introduction}

\IEEEPARstart{U}{nmanned} aerial vehicles (UAVs) are commonly used to perform field coverage activities. These vehicles are typically equipped with a large variety of sensors to take measurements of areas of interest. Applications for these kind of systems include monitoring operations in agriculture \cite{santesteban_high-resolution_2017}, archaeology \cite{themistocleous_unmanned_2015}, and civil infrastructures \cite{ham_visual_2016}. 

\begin{figure}[ht!]
\centering
\includegraphics[width=0.7\columnwidth]{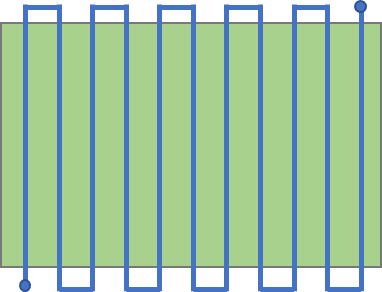}
\caption{\label{fig:boustrophedon} Example of a boustrophedon pattern used for remote sensing.}
\end{figure}

In many applications, remote sensing best practices use post-processed information in the form of an orthomosaic~\cite{colomina_unmanned_2014}. An orthomosaic is essentially a geometrically corrected image obtained thanks to the composition of several overlapped photographs~\cite{diaz-cabrera_photogrammetric_2013}. This technique implies an exhaustive and complete coverage of the area, commonly using boustrophedon patterns as the one shown in Fig.~\ref{fig:boustrophedon}. However, in the case of precision farming or other monitoring domains, the creation of a complete orthomosaic can be extremely time consuming and might require several flights to cover relatively small areas, thus limiting the real world applicability of drones. To understand the dimension of the problem, it is worth to mention the experience of the H2020 EU project PANTHEON ``Precision farming of hazelnut orchards" where the farming area consists of several hundreds of hectares while the area that can be covered for each flight using a boustrophedon approach is less than half a hectare. 

Given this common limitation in time and resources, the literature has focused on the definition of optimal policies that partially cover the area of interest. In this regard, many persistent monitoring works rely on graph-based strategies where the latency in between visits to every region is minimized~\cite{alamdari_persistent_2012,cassandras_optimal_2013,scherer_persistent_2019}. However, these strategies consider a static and node-independent distribution of the phenomena, which makes them non suitable for physical systems with significant dynamics.

Alternative approaches base their policies on the spatial correlation between the measurements. In these works, the mission path is computed so that it avoids redundant data and thus gathers the maximum amount of information per flight.  In~\cite{klesh_path_2008}, UAVs equipped with omnidirectional sensors perform an information-based exploration where the goal is to minimize the time to obtain a predefined measure of data. In environment monitoring, the planning strategy is defined such that the measurement uncertainty of a Gaussian Process (GP) regression is minimized~ \cite{gros_multi-robot_2018,Suryan2020LearningAS}. Also in the field of autonomous underwater vehicles (AUVs), multiple AUVs are used to perform the sampling of a scalar field based on the information obtained \cite{cui_mutual_2016}. These works optimize the path regarding the spatial distribution of the possible measurements. Yet, they tend to fail in the evaluation of the temporal correlation with previous data information, which is a meaningful aspect in most persistent monitoring activities.

In this paper, we assume that the phenomena to be monitored have dynamics and statistical properties which are known. This is the case of e.g. water distribution~\cite{kang_soil_nodate,skaggs_drip_2010} or dust deposition~\cite{saidan_experimental_2016,weber_performance_2014}. We believe that a good monitoring policy must take advantage of these properties in the definition of the UAV path planning. In our work, we propose a path planning strategy where the area of interest is only partially covered and the remaining elements are estimated based on the dynamics of the system and the spatial correlation between measurements. This approach resembles a sensor selection problem \cite{tzoumas_sensor_2015, joshi_sensor_2009, mo2011stochastic} where the measurement points are equivalent to the selection or not of available sensors in the observer formulation.

To the best of our knowledge, the literature presents only a few examples of path planning for spatio-temporal phenomena monitoring \cite{binney_optimizing_2013,garg_persistent_2018,lan_planning_2013}. Binney \textit{et al.} \cite{binney_optimizing_2013} define a recursive greedy algorithm to compute waypoints based on a given indicator of the estimation process. As example, given a Gaussian Process, the covariance of the estimation of different areas is minimized during a sensing exploration using AUVs. In Garg \textit{et al.} \cite{garg_persistent_2018}, the authors assume a stochastic dynamic system and perform a multi-vehicle sampling where each robot moves such that the entropy of a particle filter is maximized. Lan \textit{et al.} \cite{lan_planning_2013} model the phenomena as Gaussian processes and define periodic trajectories to minimize the largest eigenvalue of the covariance of a Kalman Filter.

In this paper, the path planning policy is obtained as part of  the estimation process of the monitored phenomena. This is achieved by structuring the path planning as an Orienteering Problem (OP) \cite{golden_orienteering_1987}. The Orienteering Problem is a combinatorial problem which consists of a node selection where the shortest path in between the selected nodes is determined. Given a time or length constraint, the objective is to maximize the score given by the visited points. We believe that UAV remote sensing activities, due to the flight time restriction and the discrete nature of the measurements, are naturally recast in OPs. In this context, the set of points represents the possible measurement coordinates, the time interval is adapted to the vehicle autonomy, and the cost function is some measure related to the measurement points.

Some works in path planning already propose tentative information-based policies based on the Orienteering problem. In~\cite{yu_correlated_2016}, the information obtained is maximized using a quadratic utility function to represent the spatial relation between the different measurement points. More recently, Bottarelli \textit{et al.}~\cite{bottarelli_orienteering-based_2019} introduce an Orienteering-based path planning used to optimize a level set estimation. In such work, the measurement points are selected such that the accuracy of the level sets classification is maximized.

The main contribution of this paper is the definition of the Orienteering Problem based on the maximization of the Fisher Information Matrix of a Kalman Filter with Intermittent observations to monitor a dynamic phenomenon. The main advantage of this approach is that the path of the UAV is computed taking into account the process dynamics, the estimation uncertainty and the existing fixed sensing structure. By doing so, it allows to define the optimal combination of the UAV remote sensing with additional sensing devices by resorting to an observer-based architecture.

This work focuses on the case where the monitored phenomena can be modelled as linear time invariant systems subject to Gaussian noise, a common assumption in field monitoring of physical phenomena~\cite{schmidt_situ_2018,fukuda_new_2004,lin_kalman_2019}. Also we assume that the monitoring is performed by a UAV with a flight time that is much shorter than the time constants of the monitored system.

The developed strategy provides an offline computation of the optimal sensing areas over one step ahead horizon of the estimation process. This approach allows obtaining the optimal path in cases where the coverage is done with unknown periodicity or when the time gap between flights is large. 

In order to solve this problem we propose a Mixed-Integer Semi-Definite Programming (MISDP) formulation where the minimum eigenvalue of the information matrix is maximized. This formulation allows to obtain the optimal solution for small instances of the problem. Additionally, for the case of large-scale scenarios, a heuristic is  proposed along with an exhaustive computational analysis.

The remainder of the paper is organized as follows. In Section~\ref{sec:problem_statement}, the problem is stated and defined. Section~\ref{sec:estimation} provides the proposed estimation measure to monitor the phenomena and Section~\ref{sec:orienteering_formulation} introduces the problem formulation for the information-based path planning. In Section~\ref{sec:heuristics}, a heuristic strategy is introduced and in Section~\ref{sec:simulation_results} simulations  are performed to compare the performance of the proposed method with traditional strategies. In Section~\ref{sec:conclusions} conclusions and future work directions are discussed.

\section{Problem Setting}\label{sec:problem_statement}

Consider a plane partitioned in $N$ areas and let the linear time-invariant system
\begin{equation}
\label{eq:state_system}
x_{k+1}
= A x_k +B u_k + w_k \\
\end{equation}
describe the dynamic phenomenon that we want to observe. We assume that the state vector is in the form
\[
x_k = \left[
\begin{array}{c}
x^1_{k} \\
...\\
x^N_{k}
\end{array}
\right]
\]
where $x^i_k \in \mathbb{R}^n$ represents the states of the system at time $k$ in the $i$th area and $u_k \in \mathbb{R}^n$ is the vector of the (measured) inputs to the system at that time instant. The complete state vector is thus $x_k \in \mathbb{R}^{Nn}$ and the matrices are $A \in \mathbb{R}^{Nn \times Nn}$ and $B \in \mathbb{R}^{Nn \times n}$. The system is subject to a process disturbance  $w_k \backsim N(0,Q)$ modeled as a stochastic Gaussian noise with covariance  $Q \in \mathbb{R}^{n N \times n N}$, which is typically non-diagonal and has nonzero terms for variables describing adjacent areas.

To estimate the state of this process, two classes of sensors are assumed available: fixed sensors and a mobile sensor. An example of this configuration can be found in water stress monitoring for precision agriculture, which typically combines fixed soil moisture sensors and periodical sensing flights carried out by UAVs.

We denote with $C_i \in \mathbb{R}^{M^i \times n}$ the measurement matrix associated to the measurements that can be potentially performed on the $i$th area. This matrix consists of two sub-matrices
\begin{equation}
    C_i = \begin{bmatrix}
 C_i^f \\
 C_i^m 
\end{bmatrix}
\quad \forall i=1, \dots, N,
\end{equation}
where $C_i^f \in \mathbb{R}^{f_i \times n}$ denotes the available fixed measurements of the $i$th area and $C_i^m \in \mathbb{R}^{m_i \times n}$ is the matrix associated to the outputs that can be measured if the area is visited by a mobile sensor.

The combination of the measurement matrices of each area provides a time-invariant observation matrix $C \in \mathbb{R}^{M \times N n}$, where $M$ is the total number of measurable states, in the form of a block-diagonal matrix,
\begin{equation}
 C=
    \begin{bmatrix}
         C_1 & \dots &  O_{M^1 \times n}\\
         \vdots & \ddots & \vdots\\
          O_{M^N \times n} & \dots & C_N\\
    \end{bmatrix},
\end{equation}
which defines the information of the system that can be accessed to through the two classes of sensors, with $M^i$ the number of states that can be retrieved from the $i$-th area.

To represent the fact that at a time instant $k$ an area might or might not be visited by the UAV, we introduce the binary variable $\gamma^i_k.$ In particular,  $\gamma^i_k=1$ if the $i$th
area is visited at time $k$, and $\gamma^i_k=0$ otherwise. Accordingly, we define the measurement selection matrix $\Gamma_k^i$ corresponding to each area as
\begin{equation}
    \Gamma_k^i=
    \begin{bmatrix}
         I_{f_i \times f_i} &   O_{f_i \times m_i}\\
         O_{m_{i,k} \times f_i} & I_{m_{i,k}\times m_{i,k}} \\
    \end{bmatrix},
\end{equation}
where $m_{i,k} = \gamma^i_k m_i.$ In other words 
\begin{equation}
    \Gamma_k^i=\left\{
    \begin{array}{lr}
    \begin{bmatrix}
         I_{f_i \times f_i} &   O_{f_i \times m_i}\\
         O_{m_{i} \times f_i} & I_{m_{i}\times m_{i}} 
    \end{bmatrix} & if \, \gamma^i_k=1  \\   
        \begin{bmatrix}
         I_{f_i \times f_i} &   O_{f_i \times m_i}
    \end{bmatrix} & if \, \gamma^i_k=0.
    \end{array}
    \right. 
\end{equation}

Accordingly, the selection of the available measurements at time $k$ is provided by the matrix $\Gamma_k \in \mathbb{R}^{M_k \times M}$, which is computed as
\begin{equation}
    \Gamma_k = \Gamma_k^1 \oplus \Gamma_k^2  \oplus  \dots \oplus \Gamma_k^N =  \begin{bmatrix}
         \Gamma_k^1 & \dots &  O_{M^1 \times n}\\
         \vdots & \ddots & \vdots\\
          O_{M^N \times n} & \dots & \Gamma_k^N\\
    \end{bmatrix},
\end{equation}
where $\oplus$ denotes the direct sum operator and $M_k$ represents the number of measurements available at time $k$.

Using this matrix we can write the measurement equation of the entire system as 
\begin{equation}
    y_k=\Gamma_k (C x_k + \nu_k),
\end{equation}
where $y_k \in \mathbb{R}^{M_k}$ is the set of measurements available to the system at time $k$ and where $\nu_k$ represents the measurement noise, which is assumed to be stochastic Gaussian noise $\nu_k \backsim N(0,R)$ with diagonal covariance matrix $R \in \mathbb{R}^{M \times M}$.

Due to autonomy limitations, at each sampling time the UAV can collect information only on a limited amount of areas. To model this, in this paper we will consider the following reasonable assumptions concerning the mobile sensor:
\begin{itemize}
\item The visit of an area is equivalent to the visit of its centroid;
\item At each time $k$ the mobile sensor has a limited maximum autonomy $T_{max,k}>0$ (e.g. in the case of a UAV this is the flight time);
\item Let the time step of the linear system be $\Delta t$, the maximum autonomy of the mobile sensor is considerably smaller, i.e. $T_{max,k} << \Delta t \enskip \forall k \in \mathbb{R}$;
\item The budget of  autonomy that is spent to go from the centroid of the area $i$ to the centroid of the area $j$ is a fixed quantity $t_{ij}$ (in the case of a UAV the time to travel from $i$ to $j$).
\end{itemize}

Accordingly, we can define the UAV trajectory as a path on an undirected and connected graph  $G = <V,E> $ where $V=\{0, 1, 2, ..., N, N+1\}$ and $E \subset V \times V,$ are the set of vertices and arcs, respectively.

The vertices $1,...,N$ represent the labels of the centroids of each area, while the vertices $0$ and $N+1$ represent pre-defined starting and ending positions for each mission (in the case of a UAV are the takeoff and the landing pads, which usually coincide). 

\begin{figure}[ht!]
\centering
\includegraphics[width=1\columnwidth]{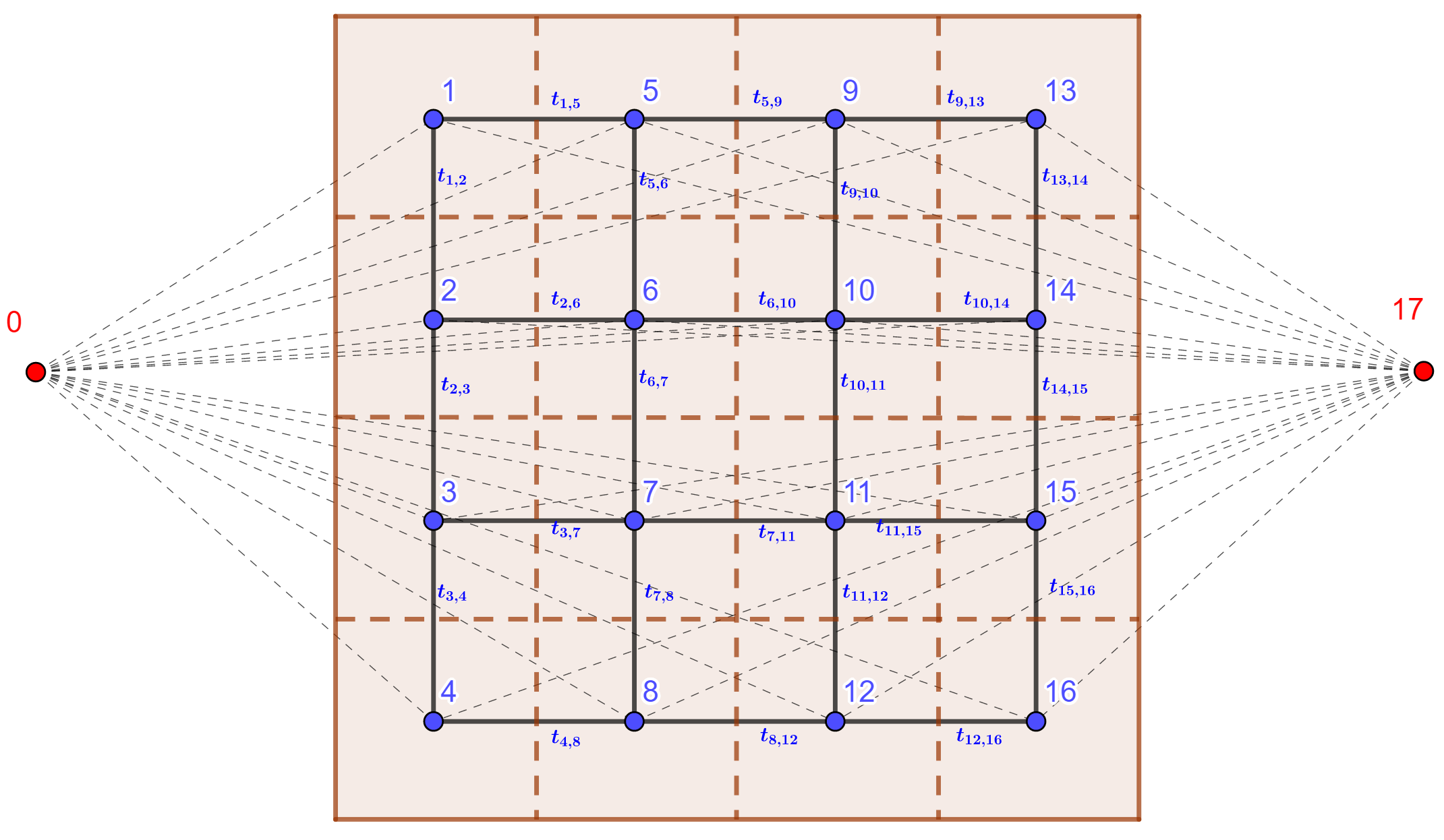}
\caption{\label{fig:Discretization} Example of the regular grid obtained for a monitored area.}
\end{figure}

Concerning the edges, in line of principle any set of arcs that makes the graph $E$ connected can be selected. In this paper, without any loss of generality,  we will focus on the realistic case that $(0,j) \in E, (j,N+1) \in E, \forall j=1,...,N$ and that an edge $(i,j)$ with $i,j\in \{1,...,N\}$ exists only if the $i$th area and the $j$th area are adjacent. This allows the remote sensing activity to start from the most convenient area $i$ and not to be constrained by the position of the takeoff and landing pads. This assumption also takes into account that in several remote sensing applications the speed during the sensing must be lower than during the takeoff and the landing. Each edge $(i,j) \in E$ has an associated weight $t_{ij}$ representing the amount of autonomy spent to travel from vertex $i$ to vertex $j$. The overall graph is depicted in Fig.~\ref{fig:Discretization}.

We reiterate that the final and initial points $0$ and $N+1$ do not need to coincide, but they may represent the same physical point depending on the application.

The goal of this paper is to compute at each sampling time $k$ a  feasible path for the UAV so that some measure of the information collected is maximized given the limited autonomy of the vehicle. In other words, this paper aims at solving the following problem

\begin{prob}
Determine the optimal ordered sequence of nodes 
\begin{equation}\label{sequence}
S_k = [s_1, s_2 , ...,s_{n_k} ]
\end{equation}
which solves the following optimization problem
\begin{eqnarray}
   & \arg\max_{S_k} f(S_k) \\
    &\text{s.t. }  s_1  =  0, \\
    & s_{n_k}  =  N+1, \\
& (s_i,s_{i+1}) \in E & i=0,...,n_k-1, \\
& \sum_{i=1}^{n_k-1} t_{s_i,s_{i+1}} \leq T_{max,k},\label{budget}
\end{eqnarray}
where $f(\cdot)$ is a suitable measure of the quality of the  collected information. In the next section we will characterize this measure of the information to be maximized.
\end{prob}

\section{Estimation of the monitored system}\label{sec:estimation}   

Since system~\eqref{eq:state_system} is a linear system subject to Gaussian noise, and given the occasional availability of the mobile sensing, the most natural choice to estimate the state is the Kalman filter with intermittent observations (see \cite{sinopoli_kalman_2004, garone_lqg_nodate}). This choice is based on the fact that the Kalman filter is the optimal estimator with  respect  to  any  quadratic  function  of the  estimation in the case of linear systems subject to  white  noise~\cite{optimal_filtering}.

The prediction step of this estimator is

\begin{equation}
   \hat{x}_{k|k-1}=A\hat{x}_{k-1|k-1}+Bu_k,
\end{equation}
\begin{equation}
    P_{k|k-1}=AP_{k-1|k-1}A^T+Q,
\end{equation}
and the correction step is
\begin{equation}
    \hat{x}_{k|k}=\hat{x}_{k|k-1}+K_{k}(y_{k}-C_{k}\hat{x}_{k|k-1}),
\end{equation}
\begin{equation}\label{eq:K_calculation}
    K_{k}=P_{k|k}C_{k}^T(C_{k}P_{k|k-1}C_{k}^T+R_{k})^{-1},
\end{equation}
\begin{equation}\label{eq:P_calculation}
    P_{k|k}=P_{k|k-1}-K_{k}C_{k}P_{k|k-1},
\end{equation}
where $\hat{x}_{k|k}$ is the estimated value of the state at time $k$ given the information available at that time and $P_{k|k}$ is the error covariance matrix of the estimation. The time-varying matrices $C_k$ and $R_k$ are defined as $C_k=\Gamma_k C$ and $R_k = \Gamma_k R \Gamma_k^T$, respectively.

The quality of the state estimation process after the $k$th data collection is typically based on the error covariance matrix $P_{k|k}.$  Combining equations~\eqref{eq:K_calculation} and \eqref{eq:P_calculation}, and applying the matrix inversion lemma, this matrix can be expressed as
\begin{equation}
P_{k|k}=[P_{k|k-1}^{-1}+C_{k}^TR_{k}^{-1}C_{k}]^{-1}.
\end{equation}
The main issue of using some function of the covariance matrix as a cost function is that, because of the inversion, this is a nonlinear function of the decision variables $\gamma_k^i$.

An alternative to the covariance matrix is the Fisher Information matrix $Y_k$. This matrix describes the quantity of information associated to each variable, and for the case of a linear system, is equivalent to $Y_{k|k}=P_{k|k}^{-1}$ ~\cite{m_segal_new_1989}. Accordingly, the post-information matrix of the estimator can be expressed as
\begin{equation}
Y_{k|k}=P_{k|k-1}^{-1}+C_{k}^TR_{k}^{-1}C_{k},
\label{eq:information_matrix}
\end{equation}
which provides a simpler expression.  Since in this paper the matrix $R_k \in \mathbb{R}^{M_k \times M_k}$ is assumed diagonal matrix, the Fisher information matrix~\eqref{eq:information_matrix} can be simplified as
\begin{equation} 
    Y_{k|k}=P_{k|k-1}^{-1}+ \sum_{i}^{M_k} \frac{C_{k,i}^T C_{k,i}}{r_{k,i}}, 
    \label{eq:lin_inform    ation_matrix}
\end{equation}
where $r_{k,i}$ is the $i$-th diagonal entry of the matrix $R_{k}$, and $C_{k,i}$ is the observation matrix when only the measurement of the $i$th entry is available. We can further simplify the expression by separating  the contribution from mobile and fixed sensors as
\begin{equation} 
    Y_{k|k}=P_{k|k-1}^{-1}+ \sum_{i=1}^N C_{f,i}^T(R_i^f)^{-1} C_{f,i} + \sum_{i=1}^N \gamma_{i,k} (C_{m,i}^T(R_i^m)^{-1} C_{m,i})
    \label{eq:lin_information_matrix}
\end{equation}
where $C_{f,i}$ represents the observation matrix $C$ whose only non-zeros entries belong to the fixed measurements of the area $i$ and, similarly, $C_{m,j}$ denotes the matrix $C$ with entries associated to the UAV measurements. In this reformulation the information matrix is conveniently defined as a linear function of the binary variable $\gamma_k^i.$

When working with covariance matrices, a common measure of performance is the trace of the covariance matrix. However, in the case of the Fisher Information matrix it has been shown \cite{grocholsky_information-theoretic_nodate} that its trace does not distinguish the gains based on the value of the eigenvalues. Therefore the trace fails to provide an appropriate measurement of the information.

In this paper, we propose as a performance objective the maximization of the minimum eigenvalue of the Fisher Information matrix. The use of the minimum eigenvalue is another way to define a \textit{measure} of a matrix, see~\cite{dette_geometry_1993}, that penalizes the uncertainty in the states with lowest Fisher information. This approach has been shown to provide a good measure of the overall uncertainty in the estimation of parameters~\cite{telen_optimal_2012,franceschini_model-based_2008}. 

The performance objective can be synthetically described as

\begin{equation}
 \begin{aligned}\label{eq:information_max}
    &\underset{\alpha, \gamma_k^i}{\text{max}} \quad \alpha\\
    &\text{s.t.} \quad \begin{multlined}[t]
    P_{k|k-1}^{-1}+ \sum_{i=1}^N C_{f,i}^T(R_i^f)^{-1} C_{f,i} +\\
    \sum_{i=1}^N \gamma_{i,k} (C_{m,i}^T(R_i^m)^{-1} C_{m,i}) \geq\alpha I. 
    \end{multlined}
    \end{aligned}
\end{equation}

In this setting, the problem becomes the one of determining at each time a sequence of nodes $S_k$ satisfying \eqref{sequence}-\eqref{budget}  that maximizes 
\eqref{eq:information_max}, where $\gamma_{i,k}=1$ if and only if $i\in S_k.$

It is very important to remark that, thanks to the statistical properties of the Linear Kalman Filter, the covariance, the information matrix and any metric associated to it, do not depend on the actual values of the measurements, but only on the covariance at time $k-1$ and of the sensing structure $\Gamma_k C$ used at time $k$.

\section{Information-based orienteering problem }\label{sec:orienteering_formulation}

The problem described in the previous two sections can be seen as a special Orienteering Problem where we have to select a subset of nodes to be visited and their order so that the information is maximized and the autonomy constraints are satisfied. 

\begin{figure*}[ht!]
  \begin{subfigure}[b]{1\columnwidth}
    \centering
    \includegraphics[width=0.8\columnwidth]{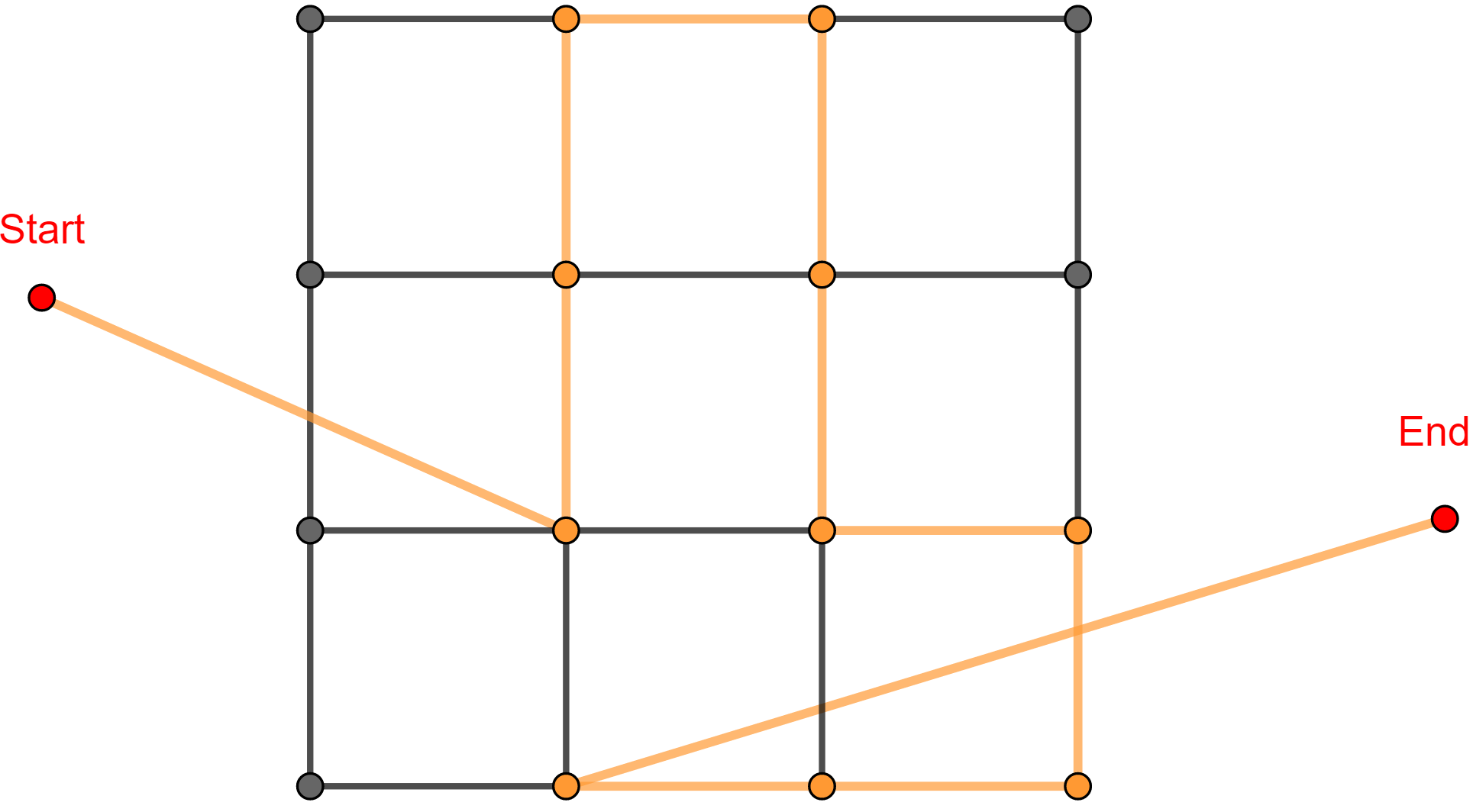} 
    \caption{Generated path 1}
  \end{subfigure}
  \begin{subfigure}[b]{1\columnwidth}
    \centering
    \includegraphics[width=0.8\columnwidth]{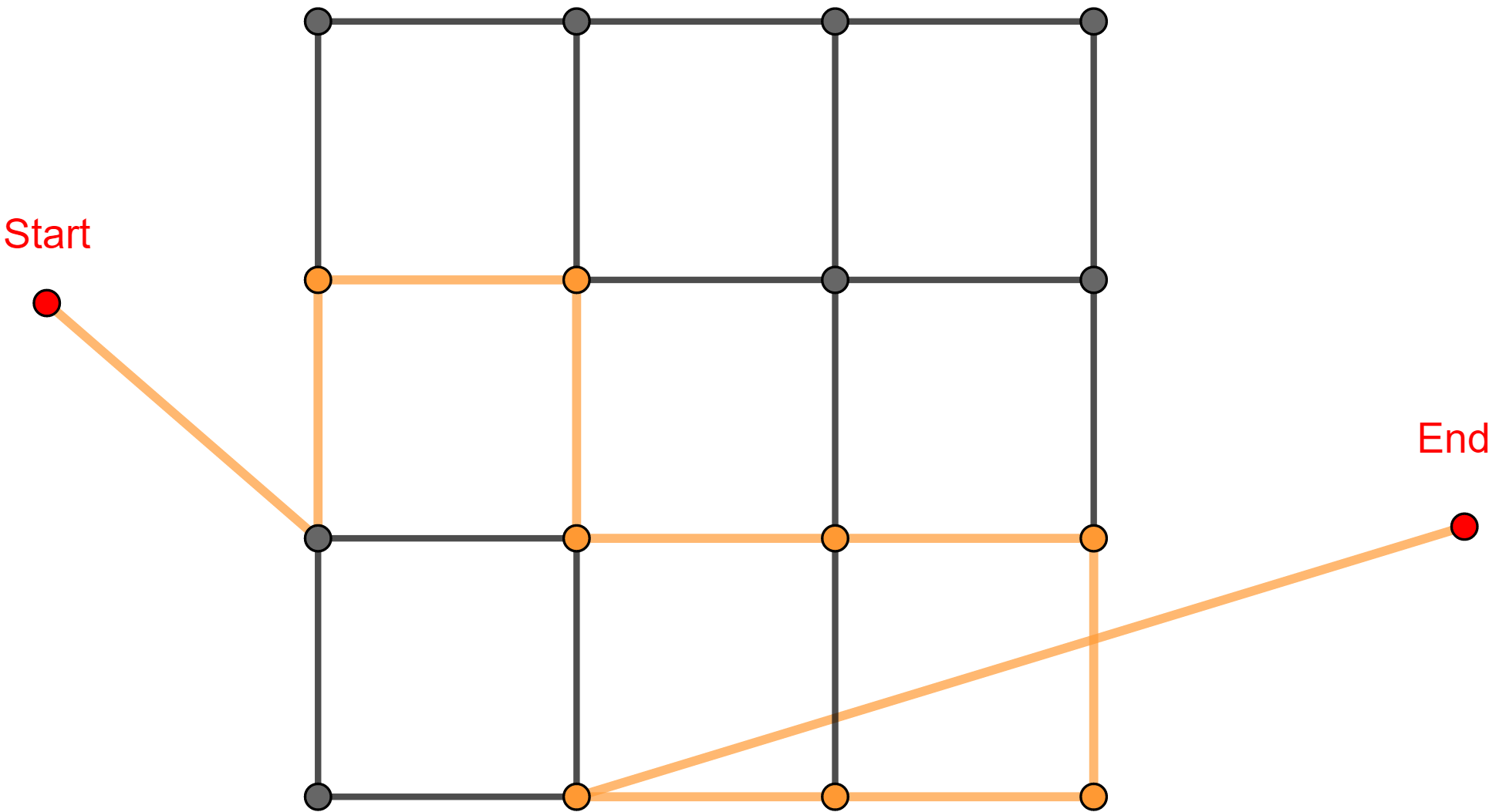} 
    \caption{Generated path 2} 
  \end{subfigure} 
  \caption{Examples of feasible paths.}
    \label{fig:feasible_path}
\end{figure*}

In this section, following an approach inspired by the Miller-Tucker-Zemlin (MTZ) formulation of the TSP \cite{miller_integer_1960}, we will propose a convenient mathematical formulation of this particular Orienteering Problem.
To this end, let us introduce two sets of decision variables; i) the binary variable $q_{ij}$  whose value is 1 if the node $j$ is visited after the node $i$ and 0 otherwise, and ii) the integer variable $u_i$ which denotes the visiting order of the node $i$.

The choice of an initial and final point is enforced by the following constraints

\begin{equation}\label{eq:forced_visit}
  \sum\limits_{i=1}^{N}q_{0i}=\sum\limits_{j=1}^{N}q_{jN+1}=1 ,
\end{equation}
\begin{equation}\label{eq:no_visit}
\sum\limits_{i=1}^{N}q_{i0}=\sum\limits_{j=1}^{N}q_{N+1j}=0 .
\end{equation}

For the rest of the nodes, we must ensure that each node is visited at most once and the continuity on the sequence of edges selected

\begin{equation}
    \sum\limits_{i \in N_p } q_{ip} = \sum\limits_{j \in N_p } q_{pj} \leq 1 ;\quad  \forall p=1, \dots, N, \label{eq:continuity}
\end{equation}
where $N_p=\{i |(i,p) \in E\}$ denotes the set of neighbors of the node $p \in V$. The endurance constraints are formalized as
\begin{equation}\label{eq:time_constraint}
\sum\limits_{i=1}^{N-1} \sum\limits_{j \in N_i} t_{ij} q_{ij} \leq T_{max,k}. \end{equation}
To avoid possible subtours and to ensure the continuity of the path, it must hold that

\begin{equation}\label{eq:subtour_1}
    2 \leq u_{i} \leq N \quad \forall i=1, \dots, N,
\end{equation}
\begin{equation}\label{eq:subtour_2}
    u_i - u_j + 1 \leq (N-1) (1-q_{ij}) \quad \forall i,j =2, \dots, N, \ (i,j) \in E.
\end{equation}

Constraints \eqref{eq:forced_visit}-\eqref{eq:subtour_2} ensure the feasibility of the UAV path. Fig.~\ref{fig:feasible_path} shows two examples of feasible and continuous paths.

Note that the fact of sensing the $i$th area at time $k$, previously introduced as $\gamma_{i,k}=1$, is equivalent to the condition $\sum_{j \in N_i} q_{ji}=1$. Therefore, combining the path planning integer constraints~\eqref{eq:forced_visit}-\eqref{eq:subtour_2}  with \eqref{eq:information_max}, where we substitute $\gamma_{i,k}=\sum_{j \in N_i} q_{ji}$, the following optimization problem is obtained
\begin{equation}\label{eq:MISDP}
\begin{aligned}
    \max_{\alpha,q,u} \quad & \alpha ,\\
    \textrm{s.t.} \enskip & 
    \begin{multlined}[t]
    P_{k|k-1}^{-1}+ \sum_{i=1}^N C_{f,i}^T(R_i^f)^{-1} C_{f,i} +\\
     \sum_{i=1}^N \sum_{j \in N_i} q_{ji} (C_{m,i}^T(R_i^m)^{-1} C_{m,i})\geq\alpha I ,\end{multlined} \\
    & \sum\limits_{i=1}^{N}q_{0i}=\sum\limits_{j=1}^{N}q_{jN+1}=1 , \\
    & \sum\limits_{i=1}^{N}q_{i0}=\sum\limits_{j=1}^{N}q_{N+1j}=0 , \\
    & \sum\limits_{i \in N_p}q_{ip} = \sum\limits_{J \in N_p}q_{pj} \leq 1 ;\quad  \forall p=1, \dots, N, \\
    & \sum\limits_{i=1}^{N-1} \sum\limits_{j \in N_i} t_{ij} q_{ij} \leq T_{max,k}, \\
    & 2 \leq u_{i} \leq N \quad \forall i=1, \dots, N, \\
    &\begin{multlined}[t]
     u_i - u_j + 1 \leq (N-1) (1-q_{ij})\\
    \quad \forall i,j =1, \dots, N+1,  (i,j), \in E,
    \end{multlined} \\
    & q_{ij} \in \{0,1\} \quad \forall i,j =0, \dots, N+1, \ (i,j) \in E. \\
\end{aligned}
\end{equation}

In this formulation, the information-based path planning is expressed as a Mixed-Integer Semidefinite Programming (MISDP) problem which, for reasonably small instances of the problem, 
can be solved optimally using commercial solvers such as  \textit{SCIP} or  \textit{cutsdp}~\cite{gally_framework_2018,c_rowe_efficient_2003}. Nevertheless, it remains a NP-hard problem whose solving time grows excessively in the case of large instances of the problem.

\section{Proposed heuristic}\label{sec:heuristics}

In this section we propose a heuristic to compute a suboptimal path based on the integer relaxation of the mixed-integer problem~\eqref{eq:MISDP}. The goal is to obtain a close-to-optimal algorithm that can be used in large-scale scenarios.

Note that the optimization problem~\eqref{eq:MISDP} becomes a convex problem when the integer variables $q_{i,j}$ and $u_i$ are relaxed. 
Consider $q_{i,j} \in [0,1]$ and $u_i \in \mathbb{R}$, such that both variables can take real values, then
Problem~\eqref{eq:MISDP} is reformulated as a Semi-Definite Programming (SDP) problem, which can be effectively and quickly solved by SDP solvers such as, e.g. \textit{Mosek} \cite{vandenberghe_semidefinite_1996}.

\begin{figure}[ht!]
\centering
\includegraphics[width=0.95\columnwidth]{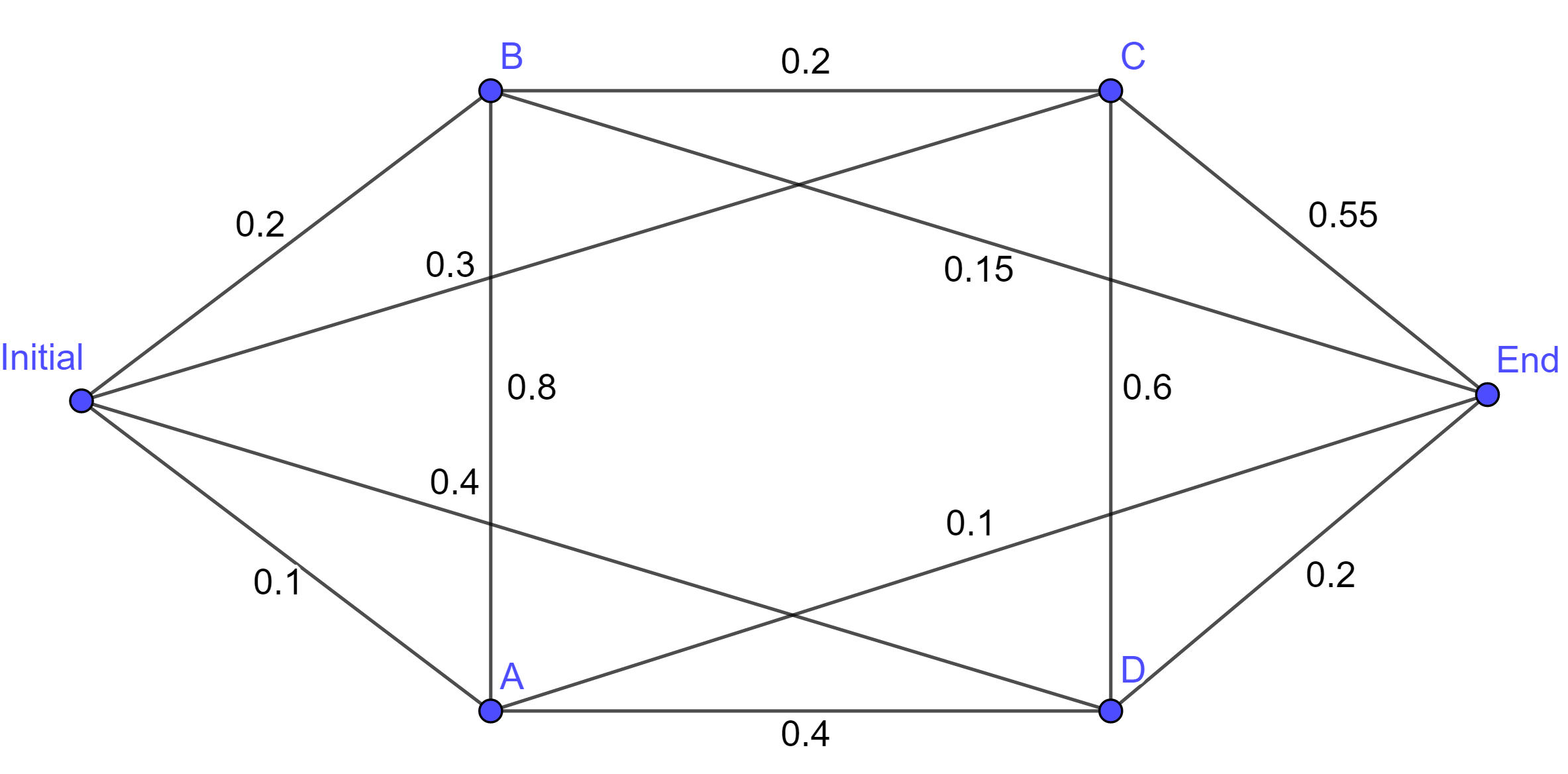}
\caption{\label{fig:relaxed_graph} Example of a possible solution of the relaxed problem.}
\end{figure}

The outcome of this convex problem, $q^r_{ij}$, provides values between $0$ and $1$ for the edges, see Fig.~\ref{fig:relaxed_graph},  which can be seen as how likely is the edge $(i,j)$ to be taken in the optimal path. Therefore, a suboptimal solution can be obtained by computing heuristics which select the points to visit based on these values~\cite{raghavan_randomized_1987}.

Algorithm \ref{Algorithm} selects the path by rounding up each link $(i,j)$ with probability $q^r_{ij}$. The algorithm starts by the initial point and sequentially adds nodes to the path based on the edges that are rounded up to 1. In this case, for each selected point $i$ there exists a set $Q_{i}^r = \cup_{j \in N_i}  q^r_{ij}$ where each element represents the probability of choosing an adjacent edge $j$ such that $\sum_{j \in N_i} q^r_{ij} = 1$. This step can be seen as a Fitness proportionate selection problem where the next edge is probabilistically chosen based on the solution of the relaxed problem. This selection can be done by using a roulette selection method or other selection-based methods~\cite{lipowski_roulette-wheel_2012,yu_roulette}.

\begin{figure*}[ht!]
  \begin{subfigure}[b]{1\columnwidth}
    \centering
    \includegraphics[width=0.9\columnwidth]{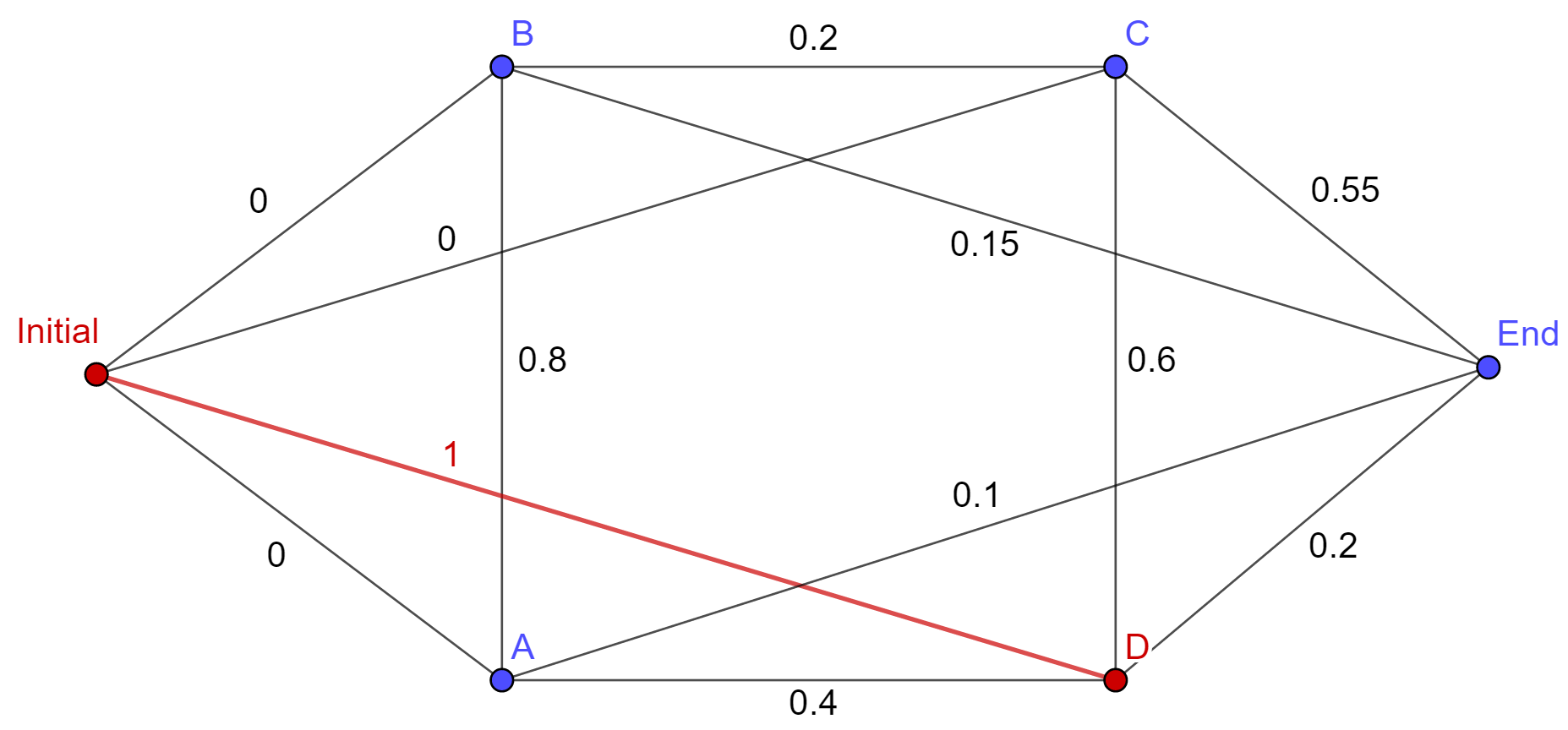} 
    \caption{Step 1}
  \end{subfigure}
  \begin{subfigure}[b]{1\columnwidth}
    \centering
    \includegraphics[width=0.9\columnwidth]{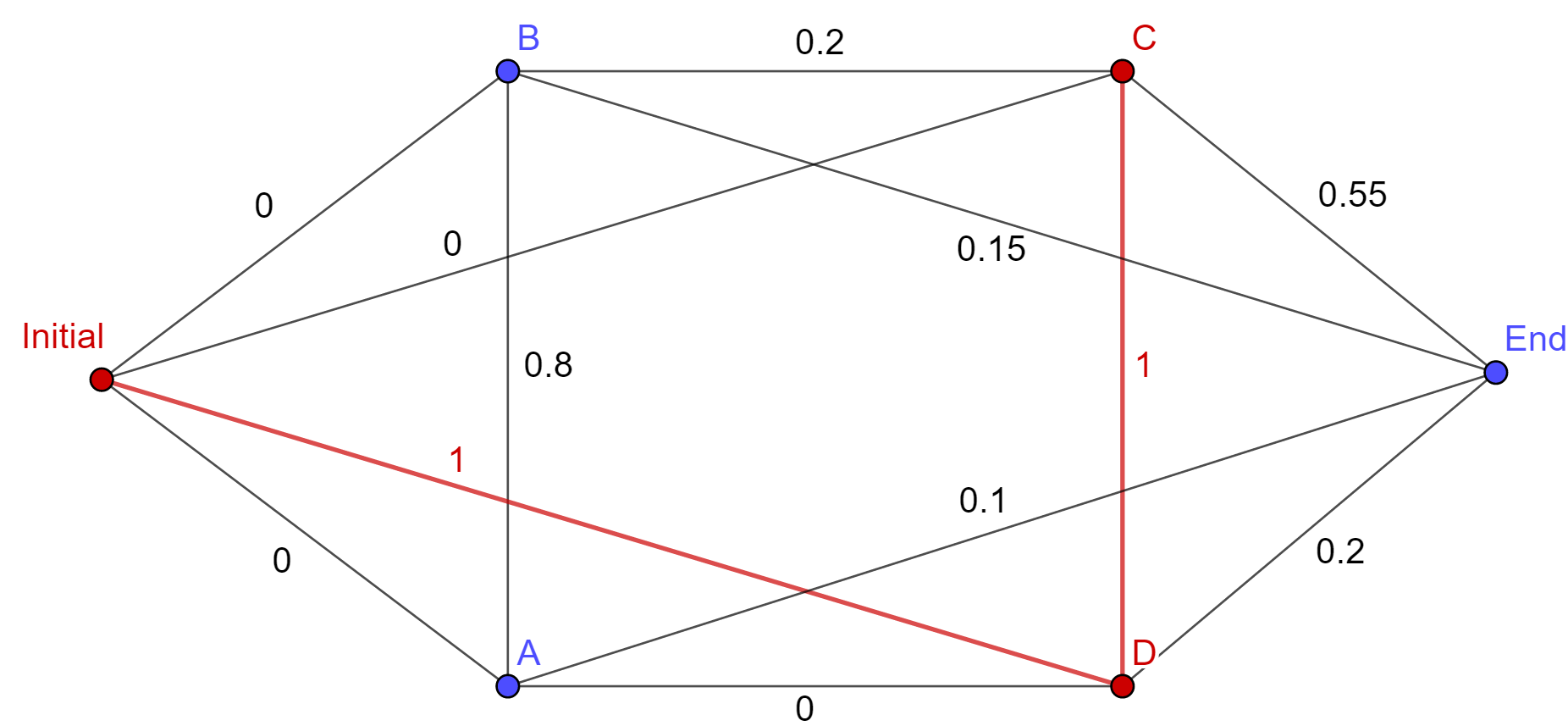} 
    \caption{Step 2} 
  \end{subfigure} 
  \caption{One step of the randomized rounding algorithm.}
    \label{fig:add_point}
\end{figure*}

To ensure that the obtained path is feasible and continuous, the rounding is done sequentially based on the previous edge selected. The algorithm starts from the initial point $V_{ini}$ adding edges until the maximum flight time $T_{max}$ is reached. This process is depicted in Fig.~\ref{fig:add_point}.

To converge to a close-to-optimal solution, the operation is repeated for a sufficiently large number of $L$ iterations, which is tuned numerically. After each iteration, the minimum eigenvalue associated to the computed path, $\lambda_1(X_{path})$, is compared with the previously stored solution. If the new path improves the currently stored sequence, the latter is replaced and the algorithm keeps seeking for alternative paths. By doing so, the computational cost is proportional to the autonomy of the vehicle (number of nodes that can be visited in one flight) and the number of iterations $L$, keeping a constant computational complexity in the rounding selection regardless of the size of the area.

\begin{algorithm}
 \caption{Sequentially edge randomized rounding}\label{Algorithm}
 \begin{algorithmic}[1]
 \renewcommand{\algorithmicrequire}{\textbf{Input:}}
 \renewcommand{\algorithmicensure}{\textbf{Output:}}
 \REQUIRE Solution $G_{rel}$
 \ENSURE  Set of edges $X_{path}$ where  $\lambda_1(X_{path})$ is maximized
   \STATE $X_{path} \leftarrow \emptyset$
   \FOR{$it \leq L $}
     \STATE $X_{temp} \leftarrow \emptyset$
     \STATE $X_{new} \leftarrow \emptyset$
     \STATE $i \leftarrow 0$ \quad \% Start from initial node $0$
     \WHILE{($T(X_{temp})+t_{i,N+1} \leq T_{max}$)}
        \STATE Select $q_{ij}$ given $(i,j) \in E$ by randomized rounding 
        \STATE $X_{new} \leftarrow q_{ij}$
        \IF{$T(X_{new})+t_{i,N+1} \leq T_{max}$)}
             \STATE $i \leftarrow j$ \quad \% Move to next node $j$
             \STATE $X_{temp} \leftarrow q_{ij}$
        \ELSE
            \STATE BREAK
        \ENDIF
     \ENDWHILE
     \STATE $X_{temp} \leftarrow q_{iN+1}$
     \IF{$\lambda_1(X_{temp})\geq \lambda_1(X_{path})$}
        \STATE $X_{path} \leftarrow X_{temp}$
     \ENDIF
 \ENDFOR
 \STATE \textbf{Output} $X_{path}$
 \end{algorithmic} 
\end{algorithm}

\section{Simulation results and application} \label{sec:simulation_results}

This section provides, through numerical simulations, an illustration of the effectiveness of the information-based path planning introduced in this paper. To do so, the adaptability of the path, the evolution over time, and the performance are analyzed and compared against a traditional covering strategy.

To provide a more constructive visualization, the simulations are based on a realistic precision agriculture scenario.

\subsection{Case study: Precision agriculture}

The effectiveness of the proposed planning method is shown through a numerical example. 
In this example, inspired by the H2020 EU project PANTHEON ``Precision farming of hazelnut orchards", we consider a hazelnut orchard
where we want  to estimate the water content of the plants and soil using the  information  collected by a drone and an agrometeorological IoT network composed of a network of fixed soil humidity sensors distributed in the orchard and a weather station providing some climate and rainfall measurements in real time.

In particular we consider an orchard of $p$ plants and $N$ soil parcels which can be described by a linear system in the form
\begin{equation} \label{eq:state_systemss}
 \begin{cases}
   x_{k+1} &= Ax_k +B u_k + B_d \hat{d}_k + w_k \\
   y_k &= \Gamma_k \left ( C x_{k} + \nu_k \right )
 \end{cases}
\end{equation}
where \mbox{$x=[x_1^T \quad x_2^T \quad x_3^T ]^T \in \mathbb{R}^{N+2p}$} is the state vector with
\mbox{$x_1=[\theta_1, \ldots, \theta_N]^T$} the soil moisture status, 
\mbox{$x_2=[W_1, \ldots, W_p]^T$} the water plant status, and 
\mbox{$x_3=[W_{rem,1}, \ldots, W_{rem,p}]^T$} the water status of the leaves. $u_k$ represents the irrigation inputs and $\hat{d}_k$ the meteorological disturbances.
The used system dynamics mimic the experimental setting proposed in the PANTHEON project, which comprises a portion of an orchard within the ``Azienda Agricola Vignola'', a farm located in the province of Viterbo, Italy. The model and the parameters used to describe the water dynamics are the ones  presented in~\cite{bono_rossello_novel_2019}. In this model it is assumed  that the fixed sensors are able to capture the value of soil moisture in the area where they are deployed and that the drone is able to measure the water status of the leaves. For further information about the model, the reader is referred to~\cite{bono_rossello_novel_2019}.

\subsection{Computational analysis of the heuristic}\label{sec:compt_analysis}

This section presents a study of the scalability, accuracy, and computational performance of the presented heuristic with respect to the optimal formulation \eqref{eq:MISDP}.

Simulations are performed given different size areas, ranging from grids of $4 \times 4$ up to $6 \times 6$ nodes. In order to obtain meaningful results, for each problem size, $100$ simulations have been performed varying the information distribution of the process. For this test, the autonomy of the vehicle is such that it allows to visit a maximum of $6$ to $9$ nodes depending on the size of the grid.

Table~\ref{tab:comp_accuracy} provides the average level of degradation provided by these simulations from the heuristic with respect to the optimal values which is computed as
\begin{equation}
    \triangle = \left(1 - \frac{\lambda_1(heuristic)}{\lambda_1(optimal)}\right) * 100.
\end{equation}

As reported in Table~\ref{tab:comp_time},  the heuristic is able to obtain results with less than $5\%$ of degradation with respect to the optimal solution, while requiring a much lower computational time.

\begin{table}[ht!]
\centering
\caption{Solution degradation between the proposed strategy and the optimal.}\label{tab:comp_accuracy}
\begin{tabular}{c c  c c} 
 \hline
  Grid size & Degradation ($\triangle$) \\ [0.5ex] 
  \hline\hline
 $4 \times 4$ & $4.80 \%$   \\
 $5 \times 5$ & $3.82 \%$  \\
 $5 \times 6$ & $3.46 \%$  \\
 \hline
\end{tabular}
\end{table}

\begin{figure}[h!]
\centering
\includegraphics[width=0.95\columnwidth,trim={1.5cm 0 1cm 0}]{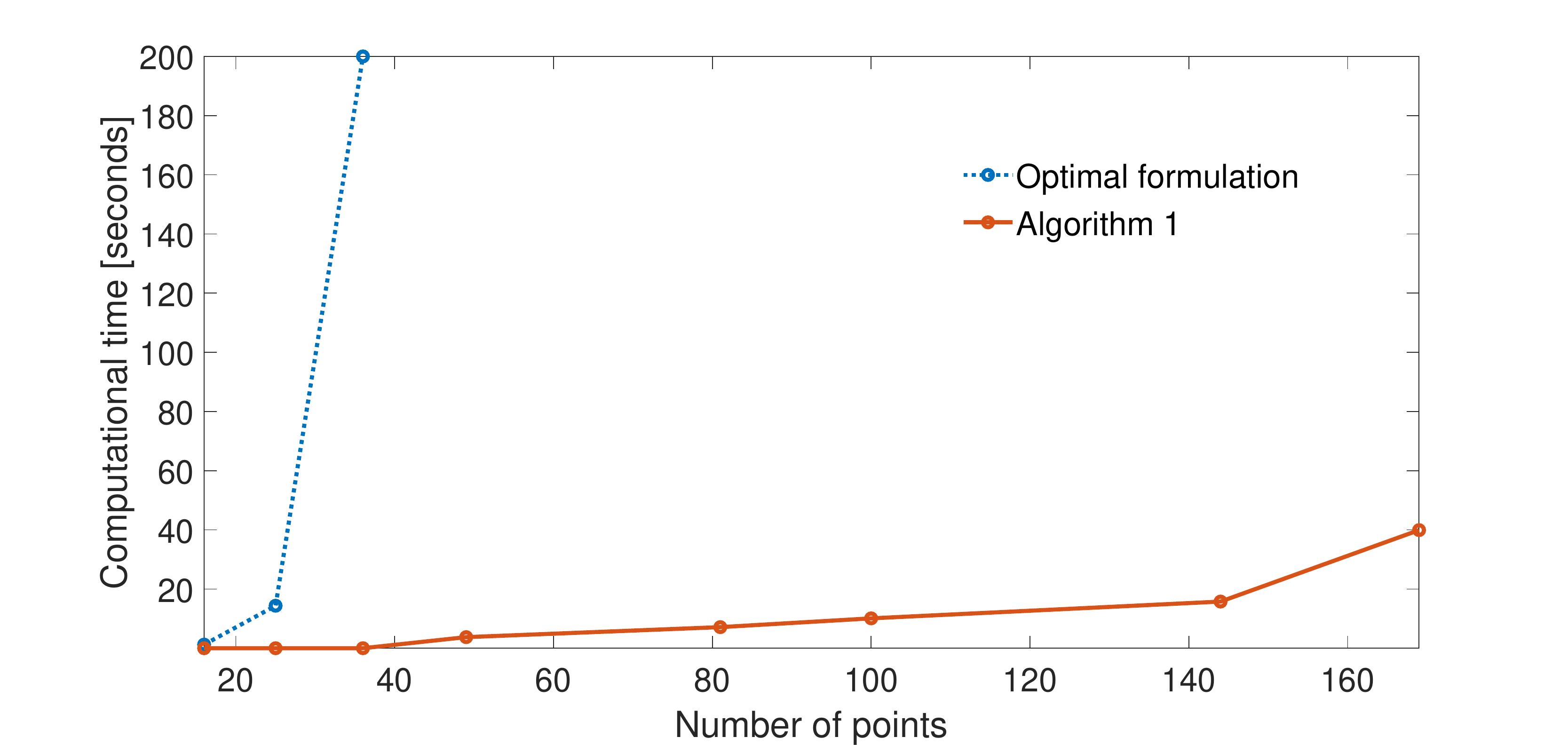}
\caption{\label{fig:compt_evol} Evolution of the computational time for the different methods.}
\end{figure}

Table~\ref{tab:comp_time} depicts the average time used to obtain the solution to the path-planning problem. In this case, for more than $36$ nodes (a grid of $6 \times 6$), we can observe how the computational time for the mixed-integer formulation of the problem becomes prohibitive. On the contrary the heuristic strategy, as shown in  Fig.~\ref{fig:compt_evol}, always provides reasonable computational times.

{ 
\renewcommand{\arraystretch}{1.2}
\begin{table}[ht!]
\centering
\caption{Comparison computational time between the optimal and the proposed strategy.}\label{tab:comp_time}
\begin{tabular}{c | c  c c c} 
\hline
\multirow{2}{*}{ Path planner} &  \multicolumn{4}{c}{Computational time (s)} \\
  & $4 \times 4$  & $5  \times 5$ & $5  \times 6$ & $6 \times 6$  \\ [0.5ex] 
  \hline\hline
 Algorithm 1  & $1.96$   & $1.91$ & $2.20$ &  $2.24$ \\
 Optimal  & $8.12$ & $169.06$ & $372.58$  & $>3600$\\
 \hline
\end{tabular}
\end{table}
}

\subsection{Simulations}

\begin{figure}[ht!]
  \begin{subfigure}[b]{0.5\columnwidth}
    \centering
    \includegraphics[width=0.75\columnwidth]{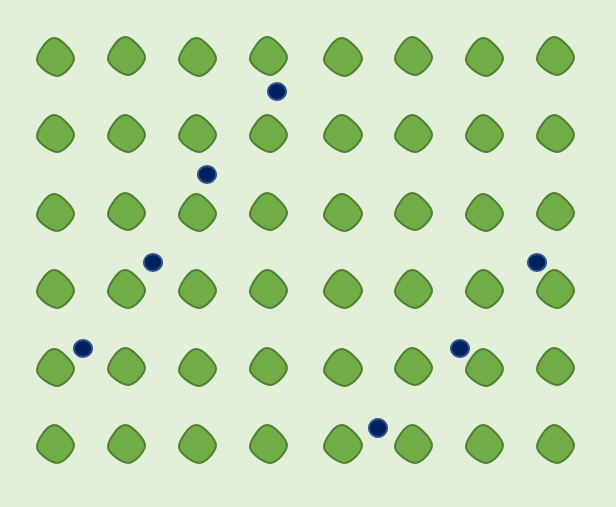} 
    \caption{Distribution 1}
  \end{subfigure}
  \begin{subfigure}[b]{0.5\columnwidth}
    \centering
    \includegraphics[width=0.75\columnwidth]{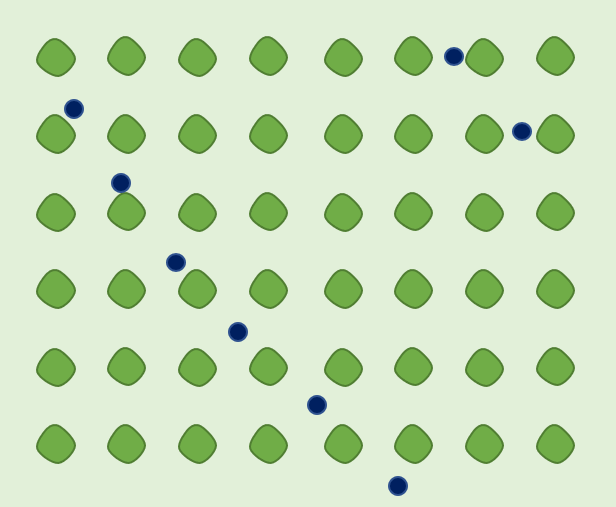} 
    \caption{Distribution 2} 
  \end{subfigure} 
  \caption{Fixed sensor distributions.}
  \label{fig:distribution} 
\end{figure}

In this section, the performance of the path planning is illustrated. Simulations are carried out for two different distributions of the ground sensors. The two scenarios are depicted in  Fig.~\ref{fig:distribution}. 
Note that the areas close to the fixed sensors have more information about the water states than the more isolated areas. Therefore one can expect that changing the location of the sensors the information distribution  changes too \cite{tzoumas_sensor_2015}, and thus the optimal covering path.

\begin{figure}[ht!]
  \begin{subfigure}[b]{0.5\columnwidth}
    \centering 
    \includegraphics[width=1\columnwidth]{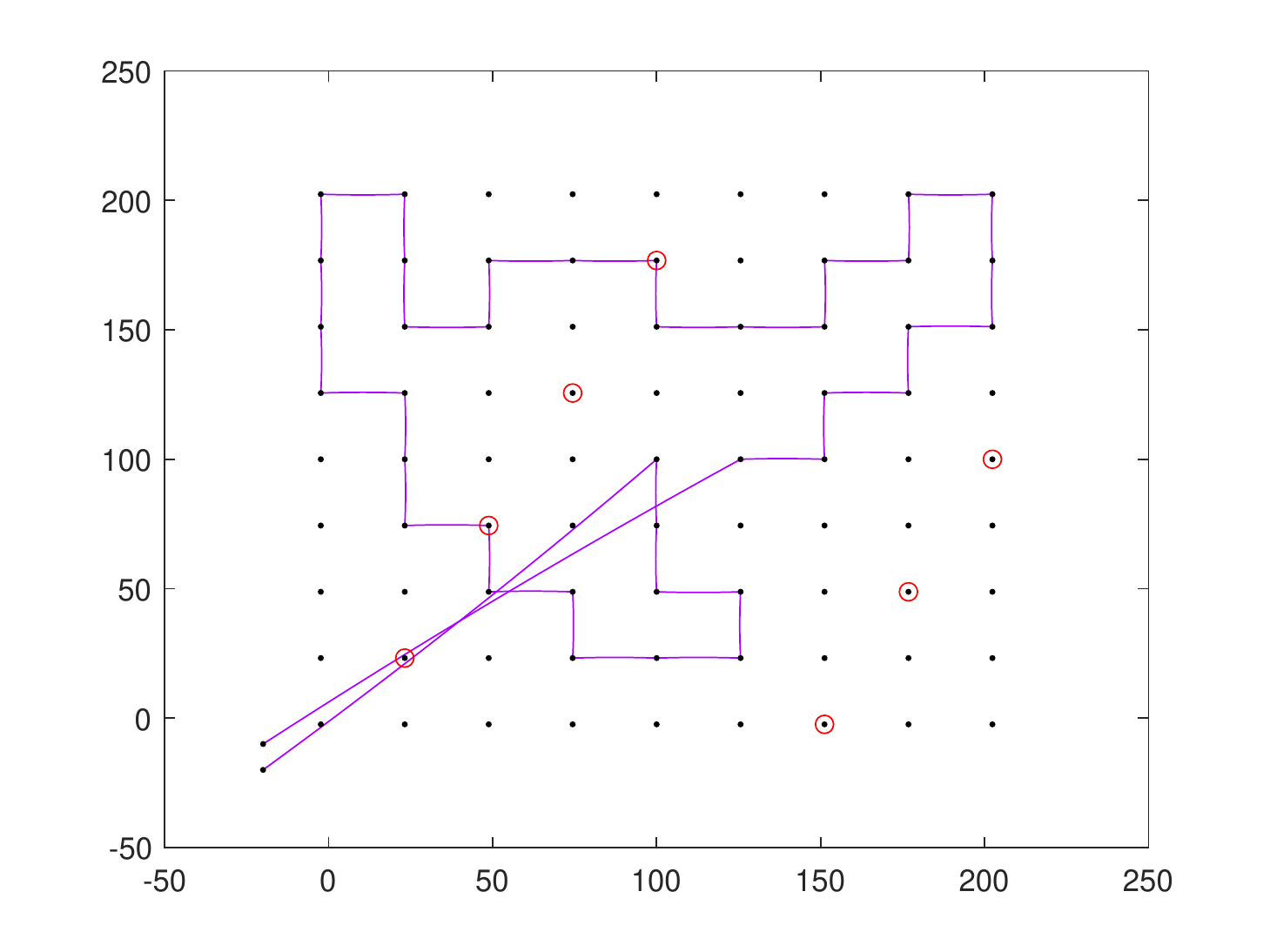} 
    \caption{Path for distribution 1} 
  \end{subfigure}
  \begin{subfigure}[b]{0.5\columnwidth}
    \centering
    \includegraphics[width=1\columnwidth]{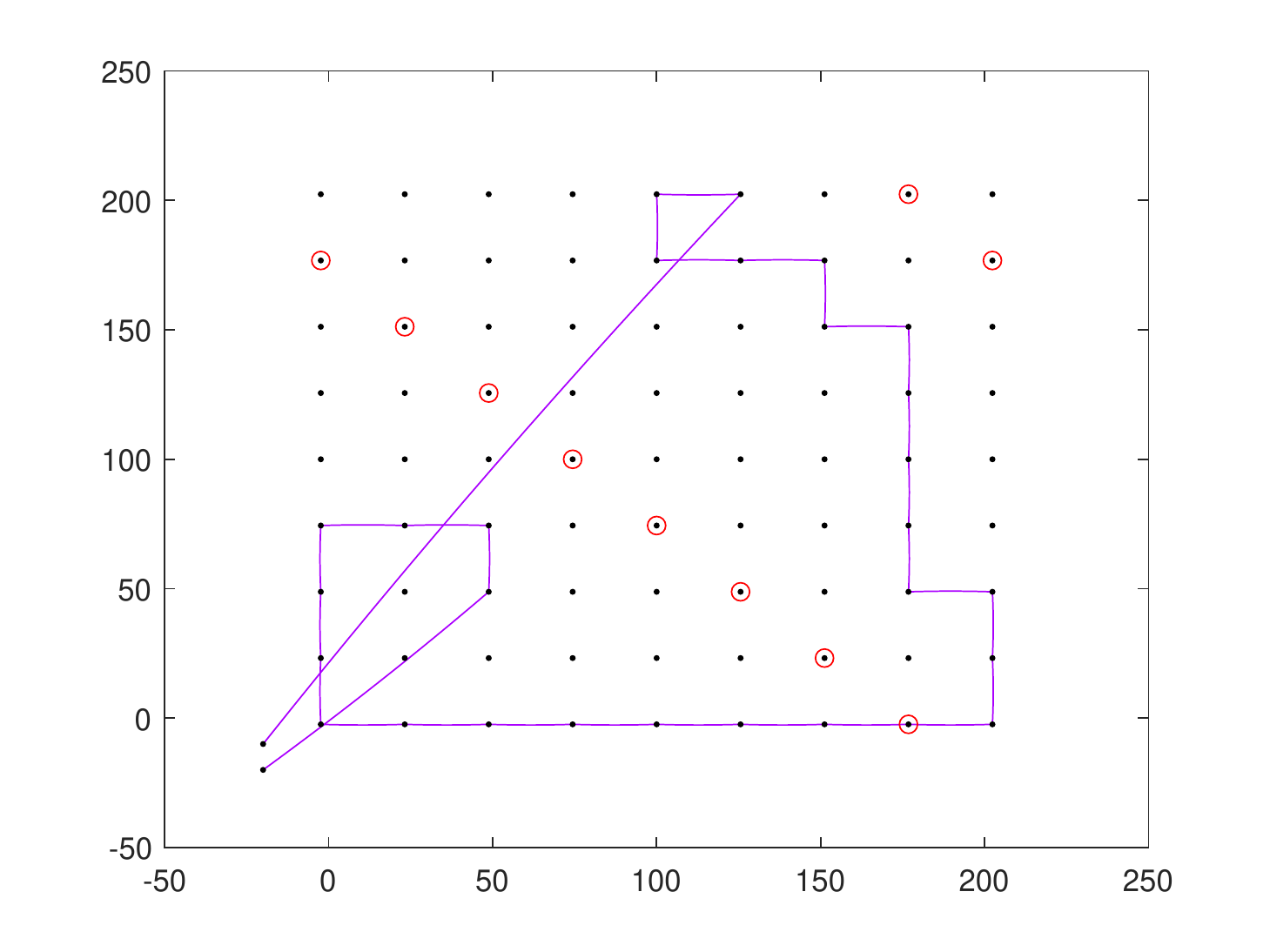} 
    \caption{Path for distribution 2} 
  \end{subfigure} 
  \caption{Comparison of the obtained path according to the soil sensor distribution.}   \label{fig:distribution_path}
\end{figure}

The paths obtained for both distributions are depicted in Fig.~\ref{fig:distribution_path}. These results show how the optimal path changes according to the fixed sensor position. It must be noted that the paths are evaluated w.r.t the information obtained and that the total path time is a constraint of the problem. Thus, a post-process reordering of the points obtained (based on TSP methods) could improve the heuristic solution in the case it allows the visit of additional points, increasing the information collected.
If the reordering does not allow to visit any additional point, as it is the case of flight (a) from Fig.~\ref{fig:distribution_path}, there might exist several feasible and optimal paths that visit the same areas in different order within the time limit.

\begin{figure}[ht] 
  \begin{subfigure}[b]{0.5\columnwidth}
    \centering
    \includegraphics[width=1.1\columnwidth]{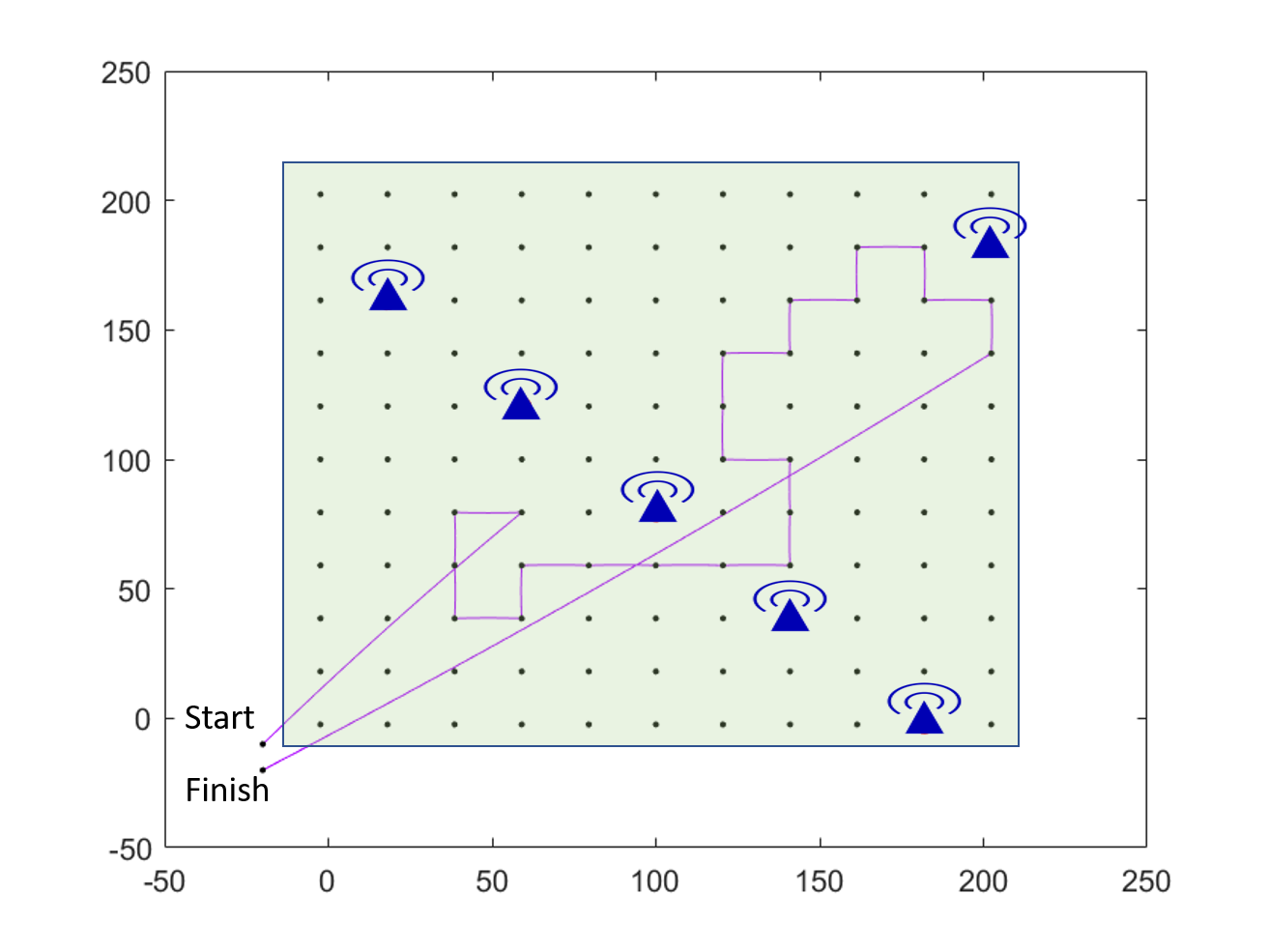} 
    \caption{Flight 1} 
    \label{fig7:a} 
    \vspace{4ex}
  \end{subfigure}
  \begin{subfigure}[b]{0.5\columnwidth}
    \centering
    \includegraphics[width=1.1\columnwidth]{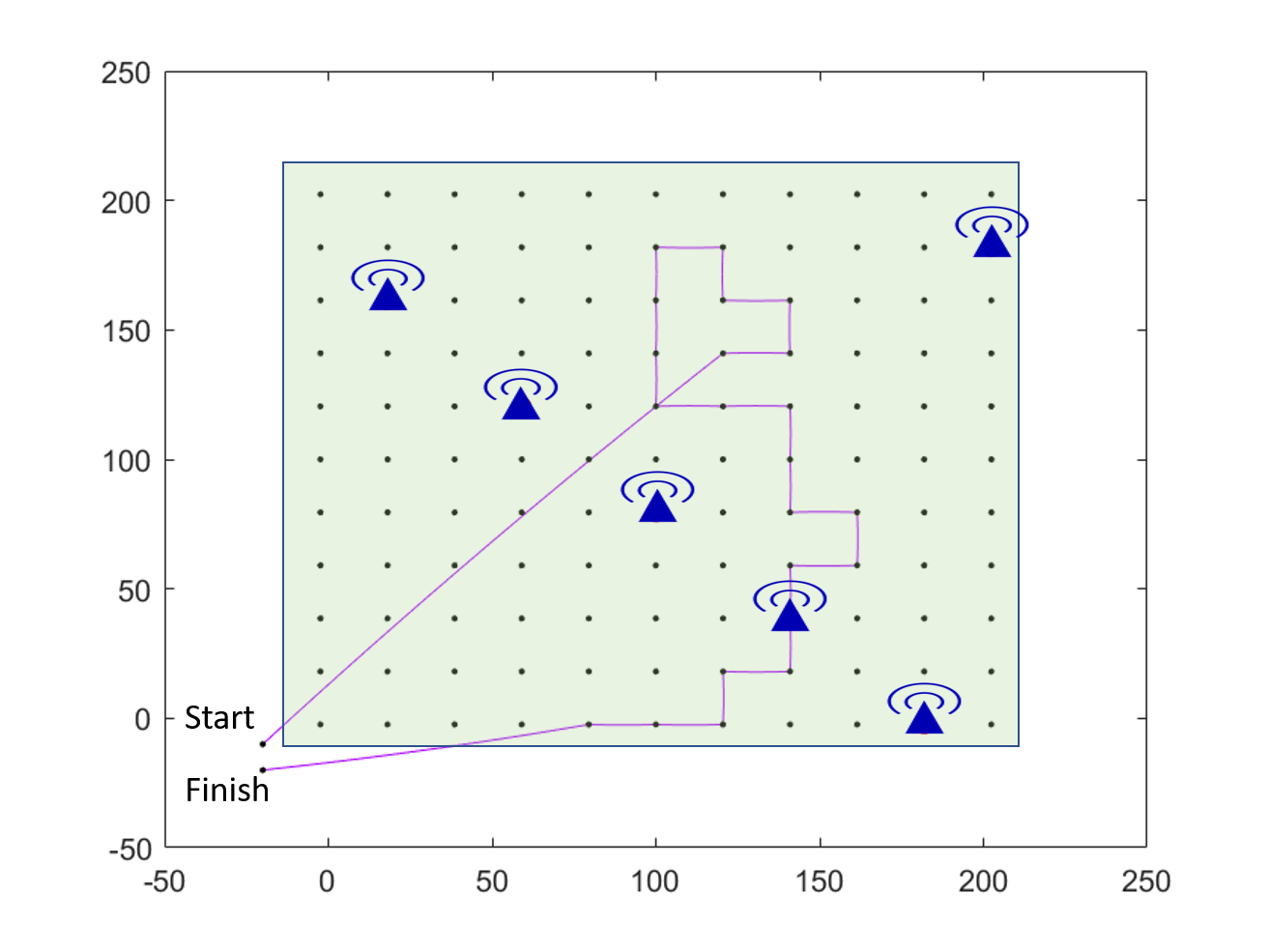} 
    \caption{Flight 2} 
    \label{fig7:b} 
    \vspace{4ex}
  \end{subfigure} 
  \begin{subfigure}[b]{0.5\columnwidth}
    \centering
    \includegraphics[width=1.1\columnwidth]{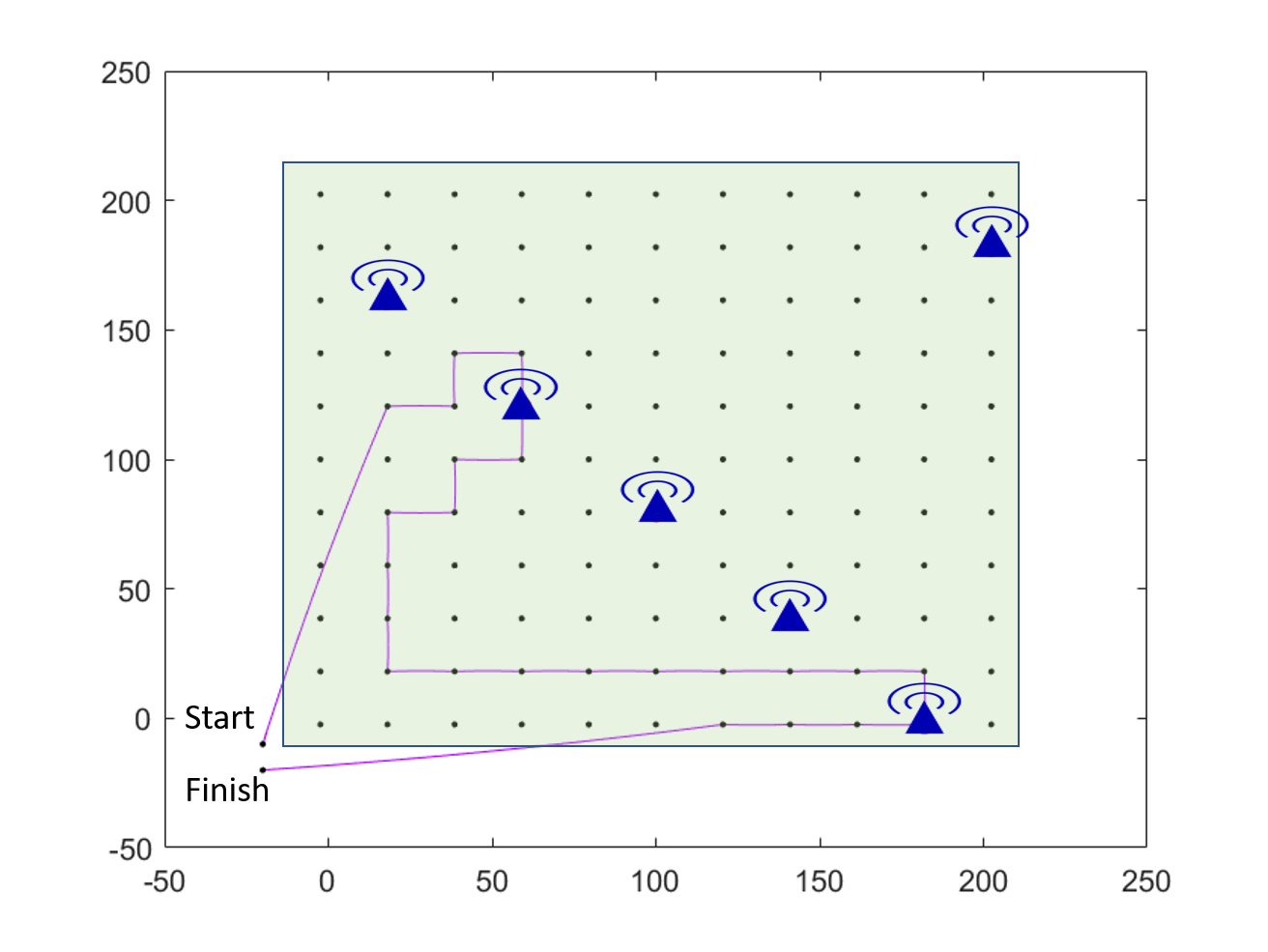} 
    \caption{Flight 3} 
    \label{fig7:c} 
  \end{subfigure}
  \begin{subfigure}[b]{0.5\columnwidth} 
    \centering
    \includegraphics[width=1.1\columnwidth]{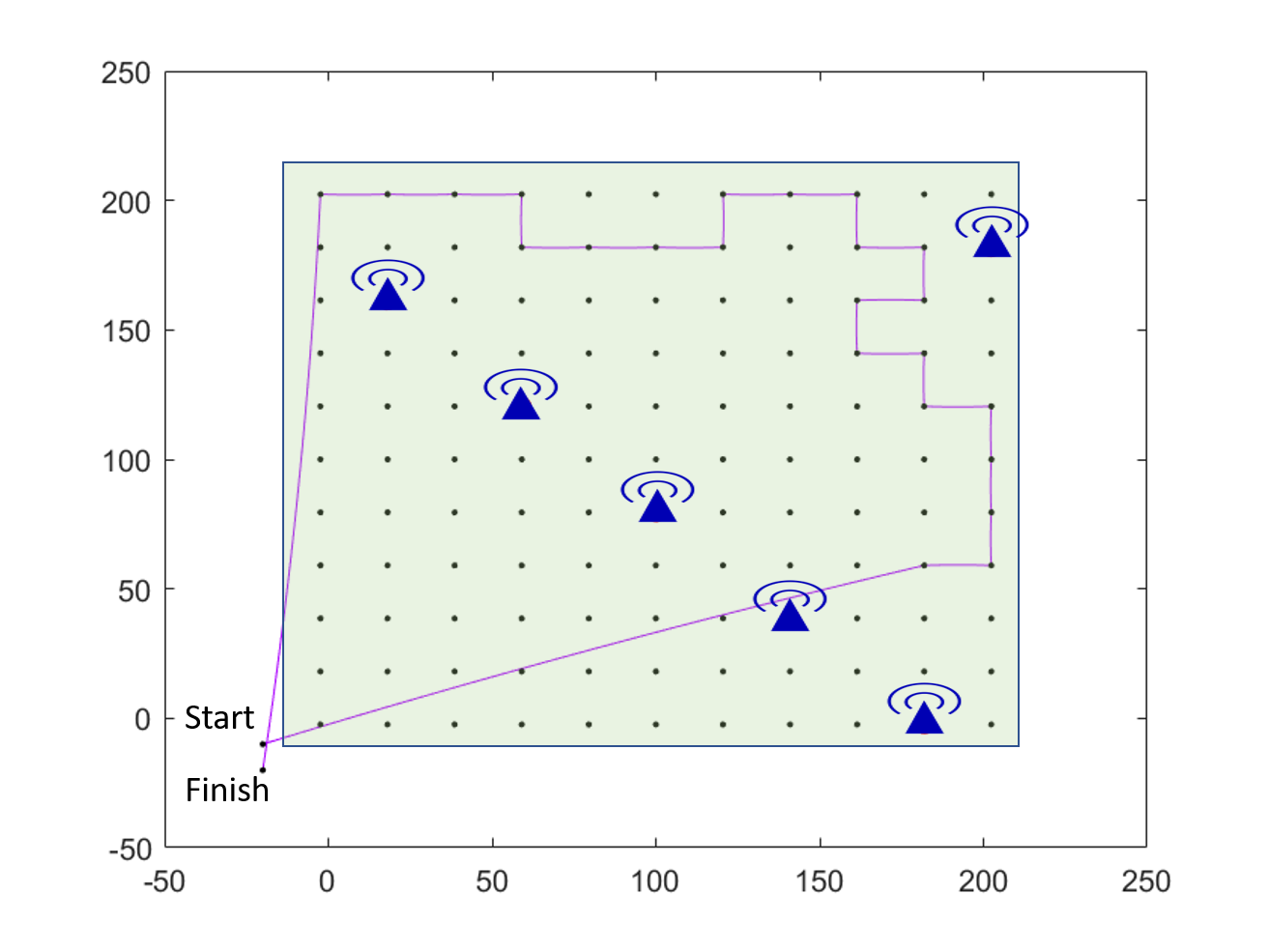} 
    \caption{Flight 4} 
    \label{fig7:d} 
  \end{subfigure} 
  \caption{Iterative path evolution.}
  \label{fig:iterative_fligths} 
\end{figure}

Clearly, the path strongly depends on the information distribution available before the measurement. To highlight this, a small area of an orchard has been simulated and four sampling periods considered. The resulting paths are depicted in Fig.~\ref{fig:iterative_fligths}. As it can be seen from the four plots, the areas with less information change accordingly to the already covered points and so does the path, demonstrating the spatial and temporal awareness of the presented strategy.

\subsection{Performance analysis} 

In this subsection, an iterative implementation of the presented path planning strategy is compared against a very common persistent monitoring strategy based on field partitioning, where the area is divided into equal number of partitions coherent with the flight autonomy and the partitions are covered sequentially~\cite{jin_optimal_2010,coombes_boustrophedon_2017}. In this case, given the symmetry of the graph and the area considered, the maximum number of points that can be visited at each flight is equal to the number of points of each partition. This fact implies that there is no sequence of flights that provides a lower latency between visits to each node. This allows comparing our strategy against a policy based only on visiting latency, another commonly used indicator in permanent monitoring~\cite{alamdari_persistent_2012,cassandras_optimal_2013}. 

To make a fair comparison, the maximum flight time obtained for the partition strategy is common for all the sensing strategies. The model is simulated over a horizon of 350 hours. The data used for the meteorological disturbances comes from the measurements of the PANTHEON weather station in the Azienda Agricola Vignola in the area of Viterbo (Italy). The flights are performed with a nonuniform sampling period ranging between $35$ and $70 $ hours, to realistically  simulate logistics and weather uncertainty. 

\begin{figure}[h!]
\centering
\includegraphics[width=1\columnwidth,trim={1.5cm 0 1cm 0}]{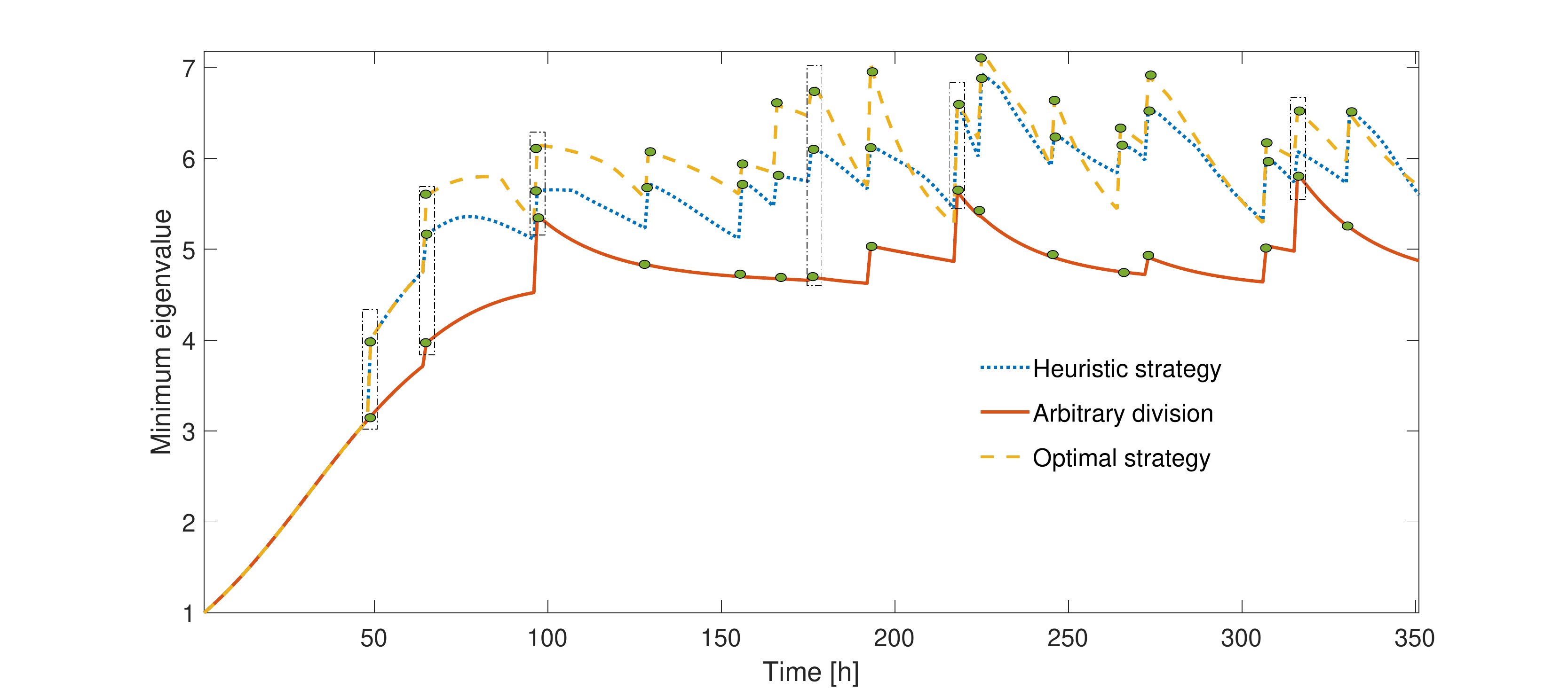}
\caption{\label{fig:comp_opt} Evolution of the minimum eigenvalue of the information matrix in a grid of $6 \times 6$ nodes. Green dots represent the time instants where a flight was performed.}
\end{figure}

First, to compare also with the optimal strategy proposed in Section~\ref{sec:orienteering_formulation}, a simulation for a  small grid of $6 \times 6$ nodes is performed. In Fig.~\ref{fig:comp_opt}, the evolution of the minimum eigenvalue of the information matrix is shown. In this figure, the partitioning strategy is referred as \textit{Arbitrary division} and \textit{Heuristic strategy} corresponds to the Algorithm~\ref{Algorithm}. From this plot, it can be seen that the presented strategies, optimal and suboptimal, clearly provide better results than the regular division. 

It is also important to notice that, in these simulations, the flights are only performed at certain time instants, and therefore the performance of the strategy must be evaluated at these points. This fact implies that the evolution of the system after the flight and therefore the initial conditions of the system at the following flights differ. This may provoke that the optimal formulation sometimes performs slightly worse than the heuristic strategy (Algorithm~\ref{Algorithm}). This is due to the fact that the combination of two independently selected optimal paths may become suboptimal in certain situations as the initial conditions for the second flight will be different. However, the results show that this may happen at a very punctual flight and that at all instants it remains clearly better than the \textit{Arbitrary division} approach. In Fig.~\ref{fig:comp_opt} some of the flight instants are indicated by a rectangle to illustrate the difference in the outcome.

\begin{figure}[h!]
\centering
\includegraphics[width=1\columnwidth,trim={1.5cm 0 1cm 0}]{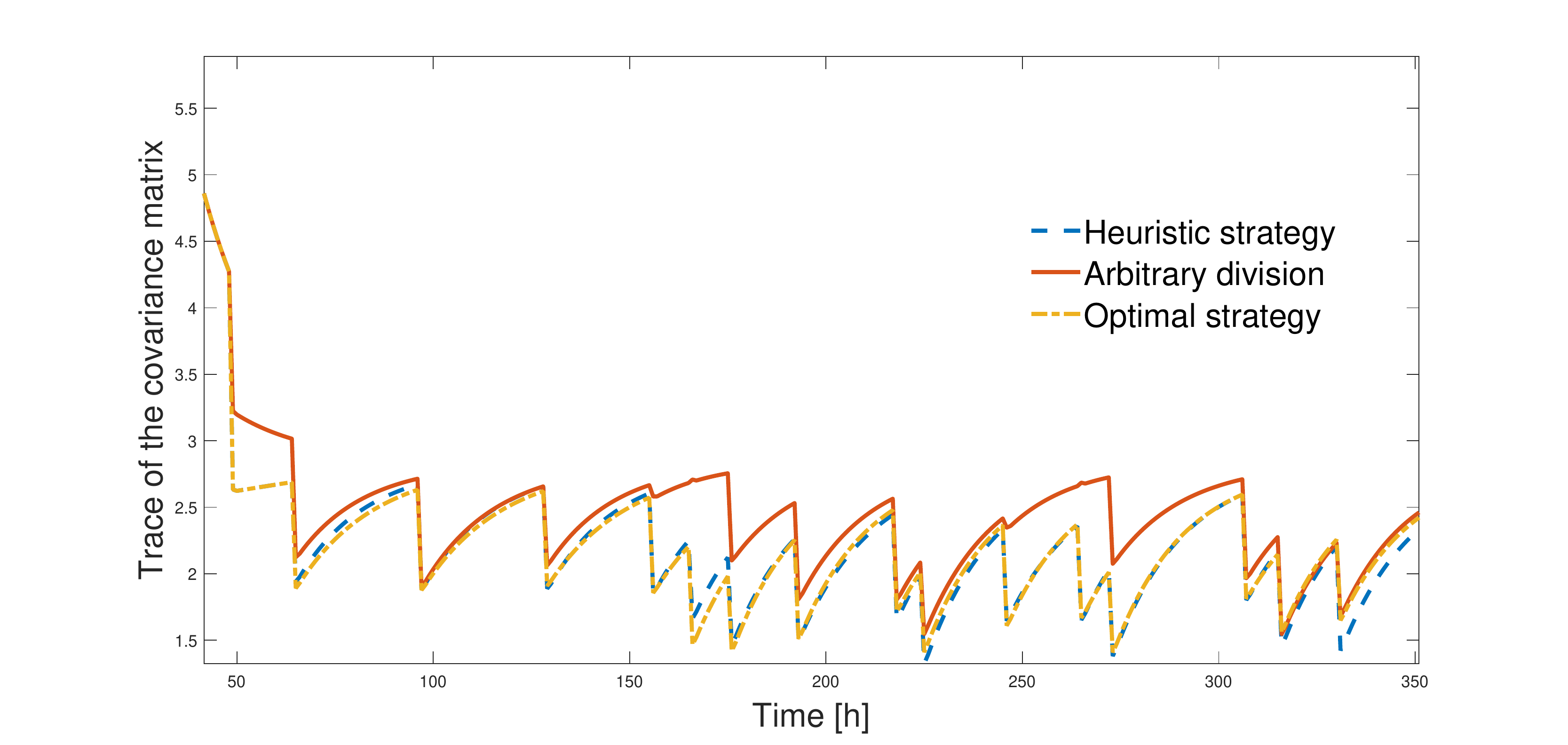}
\caption{\label{fig:trace_opt} Evolution of the trace of the covariance matrix in a grid of $6 \times 6$ nodes.}
\end{figure}

In Fig.~\ref{fig:trace_opt} the evolution of the trace of the covariance matrix is depicted. It can be seen that the presented strategy also provides much better results at steady state than the usual arbitrary strategy.

\begin{figure}[ht!]
\centering
\includegraphics[width=0.95\columnwidth]{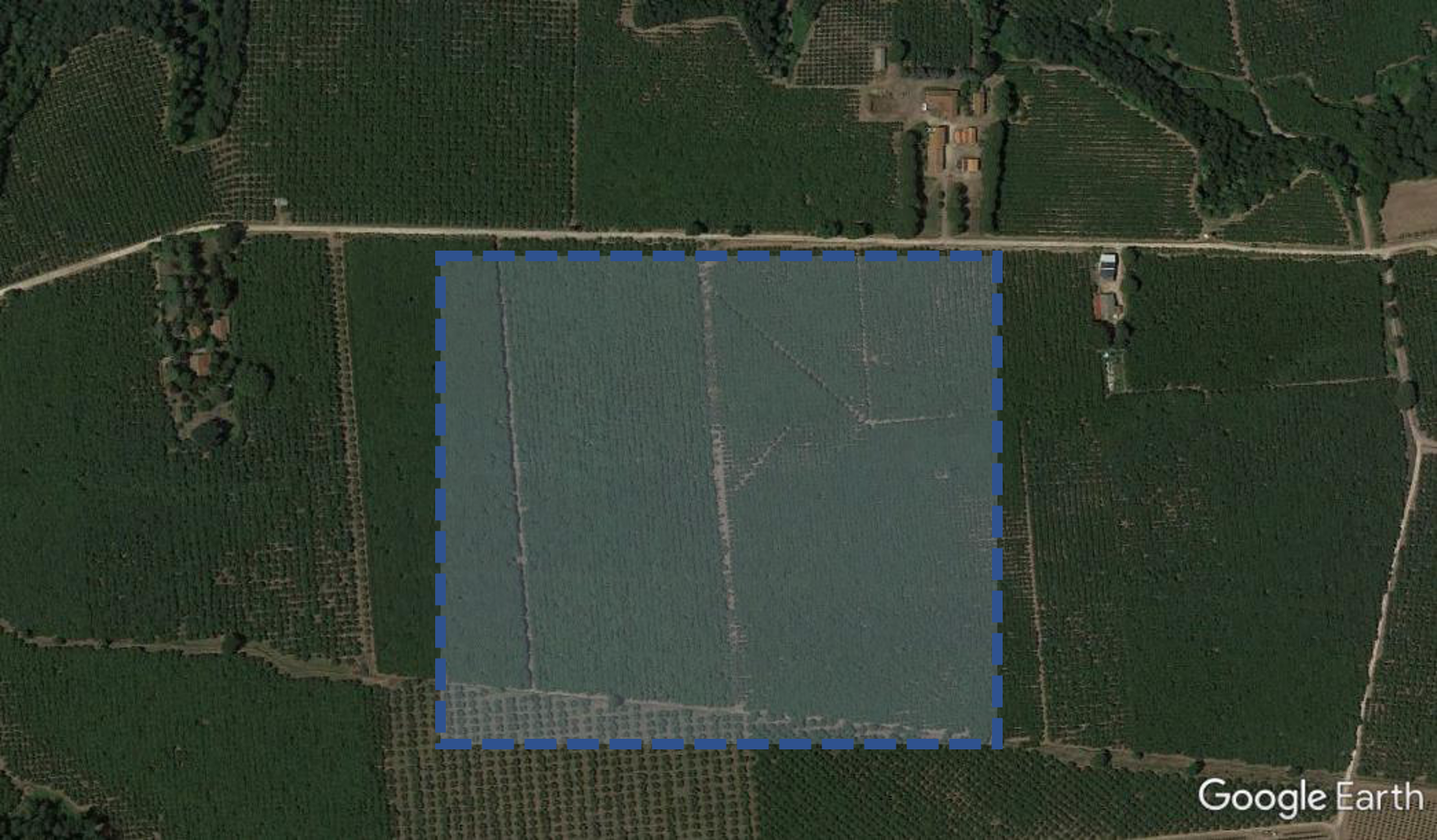}
\caption{\label{fig:image_plantation} Area considered for the simulation.}
\end{figure}

\begin{figure}[H]
\centering
\includegraphics[width=1\columnwidth,,trim={1.5cm 0 1cm 0}]{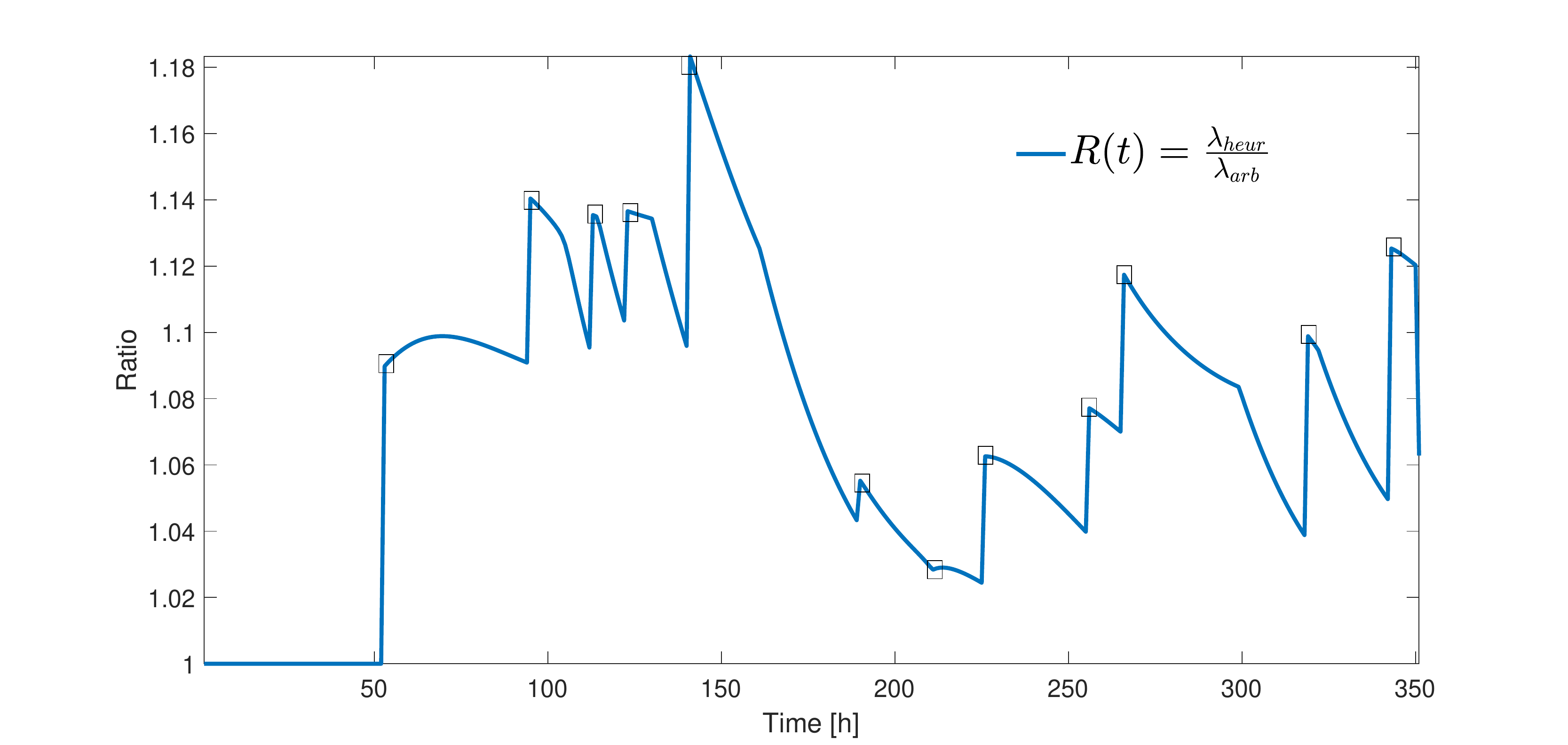}
\caption{\label{fig:comp_strategy} Ratio between the performance obtained between the strategies. Highlighted by a square the instant related to each flight.}
\end{figure}

Fig.~\ref{fig:comp_strategy} depicts the results obtained from a larger scenario, where the area dimensions correspond to a real large-scale hazelnut orchards, as shown in Fig.~\ref{fig:image_plantation}. In this case, the values represented in the plot denote the ratio between the performance of the proposed heuristic strategy and the regular division of the area, which is considered as a lower bound for our performance, and it is computed as
\begin{equation}
    R(t)=\frac{\lambda_{1,heur}(t)}{\lambda_{1,arb}(t)},
\end{equation}
where $\lambda_{1}(t)$ represents the minimum eigenvalue at time $k$ of the Fisher information matrix associated to each strategy. In such a case $R(t)>1$ indicates a better performance of the proposed strategy with respect to the regular division.

From Fig.~\ref{fig:comp_strategy}, we can observe how the heuristic strategy provides at all instants a better performance than the strategy based only on latency between visits.

\begin{figure}[H]
\centering
\includegraphics[width=1\columnwidth,,trim={1.5cm 0 1cm 0}]{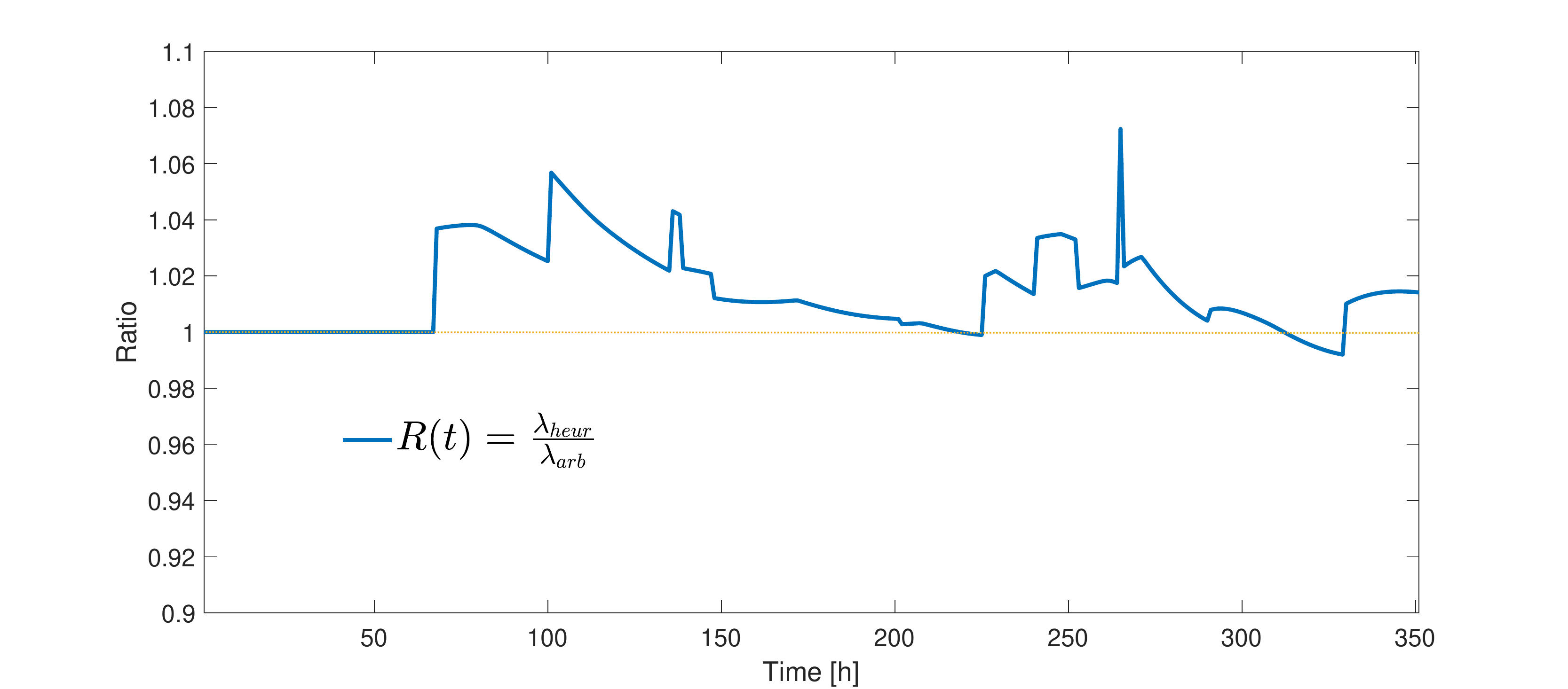}
\caption{\label{fig:comp_strategy_50} Evolution of the ratio of performance. Case with reduction of $50\%$ in flying time for the information-based approach (the flight time for the arbitrary division remains the same).}
\end{figure}

Finally, Fig.~\ref{fig:comp_strategy_50} provides, for a similar field extension and fixed sensors layout, the case where the maximum flying time for the information-based strategy (and only for it) is reduced by  $50\%$. From this plot, it can be seen that even after reducing by $50\%$ the flight time, the information-based strategy is still able to obtain really similar results, $R(t)\approx 1$ with $\pm 10\%$, to the regular coverage strategy with the double of the flight time. This result supports one of the main claims of this paper, which is that an information-based approach can help to reduce the resources put into the monitoring while keeping a similar performance. This is particularly interesting in the case of UAVs, where the main limiting factor regarding their reduced flight autonomy can be substantially mitigated by this novel strategy.

\section{Conclusion}\label{sec:conclusions}

This paper presents a novel path planning strategy. Focusing on the problem of covering  large-scale areas, we propose a path computation strategy where the flying time constraint is taken into account and the quality of the estimation of the states of the system is maximized. The problem is formulated as a special Orienteering Problem which can be written as a Mixed-Integer Semi-Definite Programming problem. Additionally, we present a heuristic with good performance.

The effectiveness of the presented strategy is shown through numerical simulations. These simulations demonstrate a clear improvement in the performance with respect to classical strategies. Moreover, the adaptability and flexibility of this approach allows to take into account the presence of fixed sensing structures and  it is a promising step for the combination with a fleet of mobile robots.

Future works will focus on adapting the presented approach to the case where several vehicles are used. Additionally, an important future line of research comprises the definition of new heuristics to solve the presented Orienteering problem in a more efficient way and/or that provide guarantees of performance with respect to the optimal solution. Other extensions will include the definition of non-myopic policies for multiple missions with fixed periodicity and the case where the time constants of the monitored system are similar to the flight time of the UAV.


%







\ifCLASSOPTIONcaptionsoff
  \newpage
\fi



%



\bibliography{Path_planning}

\begin{thebibliography}{10}

\bibitem{santesteban_high-resolution_2017}
L.~Santesteban, S.~Di~Gennaro, A.~Herrero-Langreo, C.~Miranda, J.~Royo, and
  A.~Matese, ``High-resolution {UAV}-based thermal imaging to estimate the
  instantaneous and seasonal variability of plant water status within a
  vineyard,'' {\em Agricultural Water Management}, vol.~183, pp.~49--59, Mar.
  2017.

\bibitem{themistocleous_unmanned_2015}
K.~Themistocleous, A.~Agapiou, B.~Cuca, and D.~G. Hadjimitsis, ``Unmanned
  {Aerial} {Systems} and {Spectroscopy} for {Remote} {Sensing} {Applications}
  in {Archaeology},'' {\em ISPRS - International Archives of the
  Photogrammetry, Remote Sensing and Spatial Information Sciences},
  vol.~XL-7/W3, pp.~1419--1423, Apr. 2015.

\bibitem{ham_visual_2016}
Y.~Ham, K.~K. Han, J.~J. Lin, and M.~Golparvar-Fard, ``Visual monitoring of
  civil infrastructure systems via camera-equipped {Unmanned} {Aerial}
  {Vehicles} ({UAVs}): a review of related works,'' {\em Visualization in
  Engineering}, vol.~4, Dec. 2016.

\bibitem{colomina_unmanned_2014}
I.~Colomina and P.~Molina, ``Unmanned aerial systems for photogrammetry and
  remote sensing: {A} review,'' {\em ISPRS Journal of Photogrammetry and Remote
  Sensing}, vol.~92, pp.~79--97, June 2014.

\bibitem{diaz-cabrera_photogrammetric_2013}
M.~Díaz-Cabrera, J.~Cabrera-Gámez, R.~Aguasca-Colomo, and K.~Miatliuk,
  ``Photogrammetric {Analysis} of {Images} {Acquired} by an {UAV},'' in {\em
  Computer {Aided} {Systems} {Theory} - {EUROCAST} 2013} (R.~Moreno-Díaz,
  F.~Pichler, and A.~Quesada-Arencibia, eds.), (Berlin, Heidelberg),
  pp.~109--116, Springer Berlin Heidelberg, 2013.

\bibitem{alamdari_persistent_2012}
S.~Alamdari, E.~Fata, and S.~L. Smith, ``Persistent monitoring in discrete
  environments: Minimizing the maximum weighted latency between observations,''
  {\em The International Journal of Robotics Research}, vol.~33, no.~1,
  pp.~138--154, 2014.

\bibitem{cassandras_optimal_2013}
C.~G. Cassandras, X.~Lin, and X.~Ding, ``An {Optimal} {Control} {Approach} to
  the {Multi}-{Agent} {Persistent} {Monitoring} {Problem},'' {\em IEEE
  Transactions on Automatic Control}, vol.~58, pp.~947--961, Apr. 2013.

\bibitem{scherer_persistent_2019}
J.~{Scherer} and B.~{Rinner}, ``Multi-uav surveillance with minimum information
  idleness and latency constraints,'' {\em IEEE Robotics and Automation
  Letters}, vol.~5, no.~3, pp.~4812--4819, 2020.

\bibitem{klesh_path_2008}
A.~T. Klesh, P.~T. Kabamba, and A.~R. Girard, ``Path planning for cooperative
  time-optimal information collection,'' in {\em 2008 {American} {Control}
  {Conference}}, (Seattle, WA), pp.~1991--1996, IEEE, June 2008.

\bibitem{gros_multi-robot_2018}
K.-C. Ma, Z.~Ma, L.~Liu, and G.~S. Sukhatme, ``Multi-robot {Informative} and
  {Adaptive} {Planning} for {Persistent} {Environmental} {Monitoring},'' in
  {\em Distributed {Autonomous} {Robotic} {Systems}} (R.~Groß, A.~Kolling,
  S.~Berman, E.~Frazzoli, A.~Martinoli, F.~Matsuno, and M.~Gauci, eds.),
  vol.~6, pp.~285--298, Cham: Springer International Publishing, 2018.

\bibitem{Suryan2020LearningAS}
V.~Suryan and P.~Tokekar, ``Learning a spatial field in minimum time with a
  team of robots,'' {\em IEEE Transactions on Robotics}, vol.~36,
  pp.~1562--1576, 2020.

\bibitem{cui_mutual_2016}
R.~Cui, Y.~Li, and W.~Yan, ``Mutual {Information}-{Based} {Multi}-{AUV} {Path}
  {Planning} for {Scalar} {Field} {Sampling} {Using} {Multidimensional}
  {RRT}*,'' {\em IEEE Transactions on Systems, Man, and Cybernetics: Systems},
  vol.~46, pp.~993--1004, July 2016.

\bibitem{kang_soil_nodate}
S.~Z. Kang, P.~Shi, Y.~H. Pan, Z.~S. Liang, X.~T. Hu, and J.~Zhang, ``Soil
  water distribution, uniformity and water-use efficiency under alternate
  furrow irrigation in arid areas,'' {\em Irrigation Science}, vol.~19,
  pp.~181--190, Sept. 2000.

\bibitem{skaggs_drip_2010}
T.~H. Skaggs, T.~J. Trout, and Y.~Rothfuss, ``Drip {Irrigation} {Water}
  {Distribution} {Patterns}: {Effects} of {Emitter} {Rate}, {Pulsing}, and
  {Antecedent} {Water} {Mention} of products and trade names are for the
  benefit of the reader and do not imply a guarantee or endorsement of the
  product by {USDA}.,'' {\em Soil Science Society of America Journal}, vol.~74,
  no.~6, pp.~1886--1896, 2010.
\newblock Place: Madison, WI Publisher: Soil Science Society.

\bibitem{saidan_experimental_2016}
M.~Saidan, G.~Albaali, e.~alasis, and J.~Kaldellis, ``Experimental study on the
  effect of dust deposition on solar photovoltaic panels in desert
  environment,'' {\em Renewable Energy}, vol.~92, pp.~499--505, July 2016.

\bibitem{weber_performance_2014}
B.~Weber, A.~Quiñones, R.~Almanza, and M.~D. Duran, ``Performance {Reduction}
  of {PV} {Systems} by {Dust} {Deposition},'' {\em Energy Procedia}, vol.~57,
  pp.~99--108, 2014.

\bibitem{tzoumas_sensor_2015}
V.~{Tzoumas}, A.~{Jadbabaie}, and G.~J. {Pappas}, ``Sensor placement for
  optimal kalman filtering: Fundamental limits, submodularity, and
  algorithms,'' in {\em 2016 American Control Conference (ACC)}, pp.~191--196,
  2016.

\bibitem{joshi_sensor_2009}
S.~Joshi and S.~Boyd, ``Sensor {Selection} via {Convex} {Optimization},'' {\em
  IEEE Transactions on Signal Processing}, vol.~57, pp.~451--462, Feb. 2009.

\bibitem{mo2011stochastic}
Y.~Mo, E.~Garone, A.~Casavola, and B.~Sinopoli, ``Stochastic sensor scheduling
  for energy constrained estimation in multi-hop wireless sensor networks,''
  {\em IEEE Transactions on Automatic Control}, vol.~56, no.~10,
  pp.~2489--2495, 2011.

\bibitem{binney_optimizing_2013}
J.~Binney, A.~Krause, and G.~S. Sukhatme, ``Optimizing waypoints for monitoring
  spatiotemporal phenomena,'' {\em The International Journal of Robotics
  Research}, vol.~32, pp.~873--888, July 2013.

\bibitem{garg_persistent_2018}
S.~Garg and N.~Ayanian, ``Persistent monitoring of stochastic spatio-temporal
  phenomena with a small team of robots,'' in {\em Robotics: Science and
  Systems}, (Berkeley, CA), Jul 2014.

\bibitem{lan_planning_2013}
X.~Lan and M.~Schwager, ``Planning periodic persistent monitoring trajectories
  for sensing robots in {Gaussian} {Random} {Fields},'' in {\em 2013 {IEEE}
  {International} {Conference} on {Robotics} and {Automation}}, (Karlsruhe,
  Germany), pp.~2415--2420, IEEE, May 2013.

\bibitem{golden_orienteering_1987}
B.~L. Golden, L.~Levy, and R.~Vohra, ``The orienteering problem,'' {\em Naval
  Research Logistics}, vol.~34, pp.~307--318, June 1987.

\bibitem{yu_correlated_2016}
J.~Yu, M.~Schwager, and D.~Rus, ``Correlated {Orienteering} {Problem} and its
  {Application} to {Persistent} {Monitoring} {Tasks},'' {\em IEEE Transactions
  on Robotics}, vol.~32, pp.~1106--1118, Oct. 2016.

\bibitem{bottarelli_orienteering-based_2019}
L.~Bottarelli, M.~Bicego, J.~Blum, and A.~Farinelli, ``Orienteering-based
  informative path planning for environmental monitoring,'' {\em Engineering
  Applications of Artificial Intelligence}, vol.~77, pp.~46--58, Jan. 2019.

\bibitem{schmidt_situ_2018}
F.~Schmidt, H.~M. Wainwright, B.~Faybishenko, M.~Denham, and C.~Eddy-Dilek,
  ``\textit{{In} {Situ}} {Monitoring} of {Groundwater} {Contamination} {Using}
  the {Kalman} {Filter},'' {\em Environmental Science \& Technology}, vol.~52,
  pp.~7418--7425, July 2018.

\bibitem{fukuda_new_2004}
J.~Fukuda, T.~Higuchi, S.~Miyazaki, and T.~Kato, ``A new approach to
  time-dependent inversion of geodetic data using a {Monte} {Carlo} mixture
  {Kalman} filter,'' {\em Geophysical Journal International}, vol.~159,
  pp.~17--39, Oct. 2004.

\bibitem{lin_kalman_2019}
Z.~Lin, H.~H.~T. Liu, and M.~Wotton, ``Kalman {Filter}-{Based} {Large}-{Scale}
  {Wildfire} {Monitoring} {With} a {System} of {UAVs},'' {\em IEEE Transactions
  on Industrial Electronics}, vol.~66, pp.~606--615, Jan. 2019.

\bibitem{sinopoli_kalman_2004}
B.~Sinopoli, L.~Schenato, M.~Franceschetti, K.~Poolla, M.~Jordan, and
  S.~Sastry, ``Kalman {Filtering} {With} {Intermittent} {Observations},'' {\em
  IEEE Transactions on Automatic Control}, vol.~49, pp.~1453--1464, Sept. 2004.

\bibitem{garone_lqg_nodate}
E.~Garone, B.~Sinopoli, A.~J. Goldsmith, and A.~Casavola, ``Lqg control for
  mimo systems over multiple erasure channels with perfect acknowledgment,''
  {\em IEEE Transactions on Automatic Control}, vol.~57, pp.~450--456, 2012.

\bibitem{optimal_filtering}
``Optimal filtering brian d. o. anderson, john b. moore,'' 1979.

\bibitem{m_segal_new_1989}
{M. Segal} and {E. Weinstein}, ``A new method for evaluating the log-likelihood
  gradient, the {Hessian}, and the {Fisher} information matrix for linear
  dynamic systems,'' {\em IEEE Transactions on Information Theory}, vol.~35,
  pp.~682--687, May 1989.

\bibitem{grocholsky_information-theoretic_nodate}
B.~{Grocholsky}, A.~{Makarenko}, and H.~{Durrant-Whyte},
  ``Information-theoretic coordinated control of multiple sensor platforms,''
  in {\em 2003 IEEE International Conference on Robotics and Automation (Cat.
  No.03CH37422)}, vol.~1, pp.~1521--1526 vol.1, 2003.

\bibitem{dette_geometry_1993}
H.~Dette and W.~J. Studden, ``Geometry of {E}-{Optimality},'' {\em The Annals
  of Statistics}, vol.~21, pp.~416--433, Mar. 1993.

\bibitem{telen_optimal_2012}
D.~Telen, F.~Logist, E.~Van~Derlinden, I.~Tack, and J.~Van~Impe, ``Optimal
  experiment design for dynamic bioprocesses: {A} multi-objective approach,''
  {\em Chemical Engineering Science}, vol.~78, pp.~82--97, Aug. 2012.

\bibitem{franceschini_model-based_2008}
G.~Franceschini and S.~Macchietto, ``Model-based design of experiments for
  parameter precision: {State} of the art,'' {\em Chemical Engineering
  Science}, vol.~63, pp.~4846--4872, Oct. 2008.

\bibitem{miller_integer_1960}
C.~E. Miller, A.~W. Tucker, and R.~A. Zemlin, ``Integer {Programming}
  {Formulation} of {Traveling} {Salesman} {Problems},'' {\em Journal of the
  ACM}, vol.~7, pp.~326--329, Oct. 1960.

\bibitem{gally_framework_2018}
T.~Gally, M.~E. Pfetsch, and S.~Ulbrich, ``A framework for solving
  mixed-integer semidefinite programs,'' {\em Optimization Methods and
  Software}, vol.~33, pp.~594--632, May 2018.
\newblock Publisher: Taylor \& Francis.

\bibitem{c_rowe_efficient_2003}
{C. Rowe} and {J. Maciejowski}, ``An efficient algorithm for mixed integer
  semidefinite optimisation,'' in {\em Proceedings of the 2003 {American}
  {Control} {Conference}, 2003.}, vol.~6, pp.~4730--4735 vol.6, June 2003.
\newblock Journal Abbreviation: Proceedings of the 2003 American Control
  Conference, 2003.

\bibitem{vandenberghe_semidefinite_1996}
L.~Vandenberghe and S.~Boyd, ``Semidefinite {Programming},'' {\em SIAM Review},
  vol.~38, pp.~49--95, Mar. 1996.
\newblock Publisher: Society for Industrial and Applied Mathematics.

\bibitem{raghavan_randomized_1987}
P.~Raghavan and C.~D. Tompson, ``Randomized rounding: {A} technique for
  provably good algorithms and algorithmic proofs,'' {\em Combinatorica},
  vol.~7, pp.~365--374, Dec. 1987.

\bibitem{lipowski_roulette-wheel_2012}
A.~Lipowski and D.~Lipowska, ``Roulette-wheel selection via stochastic
  acceptance,'' {\em Physica A: Statistical Mechanics and its Applications},
  vol.~391, pp.~2193--2196, Mar. 2012.

\bibitem{yu_roulette}
F.~{Yu}, X.~{Fu}, H.~{Li}, and G.~{Dong}, ``Improved roulette wheel
  selection-based genetic algorithm for tsp,'' in {\em 2016 International
  Conference on Network and Information Systems for Computers (ICNISC)},
  pp.~151--154, 2016.

\bibitem{bono_rossello_novel_2019}
N.~Bono~Rossello, R.~Fabrizio~Carpio, A.~Gasparri, and E.~Garone, ``A novel
  {Observer}-based {Architecture} for {Water} {Management} in {Large}-{Scale}
  ({Hazelnut}) {Orchards},'' {\em IFAC-PapersOnLine}, vol.~52, no.~30,
  pp.~62--69, 2019.

\bibitem{jin_optimal_2010}
{J. Jin} and {L. Tang}, ``Optimal {Coverage} {Path} {Planning} for {Arable}
  {Farming} on {2D} {Surfaces},'' {\em Transactions of the ASABE}, vol.~53,
  no.~1, pp.~283--295, 2010.

\bibitem{coombes_boustrophedon_2017}
M.~Coombes, W.-H. Chen, and C.~Liu, ``Boustrophedon coverage path planning for
  {UAV} aerial surveys in wind,'' in {\em 2017 {International} {Conference} on
  {Unmanned} {Aircraft} {Systems} ({ICUAS})}, (Miami, FL, USA), pp.~1563--1571,
  IEEE, June 2017.

\end{thebibliography}
\bibliographystyle{ieeetr}

%









\end{document}